\documentclass[pra,amsmath,amssymb,twocolumn,superscriptaddress,showpacs]{revtex4-2}

\usepackage[utf8]{inputenc}
\usepackage{graphicx}
\usepackage{dcolumn}
\usepackage{bm}
\usepackage{hyperref}
\usepackage{soul}

\begin{document}
	
	\preprint{APS/123-QED}
	
	\title{Rotating-modulated Higher-Order Topological States in a Split-ring Photonic Insulator}
	
	\author{Hui Chang Li}
	\affiliation{School of Physics and Materials Science, Nanchang University, Nanchang 330031, China;}
	
	\author{Xiang Zhou}%
	\affiliation{School of Physics and Materials Science, Nanchang University, Nanchang 330031, China;}
	
	\author{Hai Lin Chi}
	\affiliation{School of Physics and Materials Science, Nanchang University, Nanchang 330031, China;}
	
	\author{Wen Wen Wang}%
	\affiliation{School of Physics and Materials Science, Nanchang University, Nanchang 330031, China;}%
	
	\author{Yun Shen}%
	\email{shenyun@ncu.edu.cn}
	\affiliation{School of Physics and Materials Science, Nanchang University, Nanchang 330031, China;}%
	\affiliation{Institute of Space Science and Technology, Nanchang University, Nanchang 330031, China}%
	
	\author{Xiao Hua Deng}%
	\email{dengxiaohua0@gmail.com}
	\affiliation{School of Physics and Materials Science, Nanchang University, Nanchang 330031, China;}%
	\affiliation{Institute of Space Science and Technology, Nanchang University, Nanchang 330031, China}%
	
	\date{\today}
	
	\begin{abstract}
		The emerging field of topology has brought device effects to a new level. Higher-order topological insulators (HOTIs) go beyond traditional descriptions of bulk-edge correspondence, broadening the understanding of topologically insulating phases. In this paper, a second-order split-ring photonic crystal (SSPC) with zero-dimensional ($0$D) corner states and one-dimensional ($1$D) edge states is proposed. Based on the coupling strength determined by the opening direction between the split-rings, the electronic transition strength of the electronic system is imitated, and the topological trivial and non-trivial transformation of the topological two-dimensional ($2$D) SSH model are realized by using the rotating split-ring lattice. Theory and simulation find that SSPC has non-trivial topological edge states that can be quantified by bulk polarization. As the opening direction of the split-rings gradually changes within one period, there will be transitions between four different topological polarizations of the lowest energy bands, which can be conveniently used to achieve transitions between different topological phases. Our research can be extended to higher dimensions and broaden research paths for higher-order photonic topological insulators and semimetals.
	\end{abstract}
	
	\maketitle
	
	\section{\label{sec:level1}Introduction}
	Recent years, the research on topological insulators in condensed matter has flourished. Researchers have conducted extensive and in-depth research on colorful topological physical effects, from the initial quantum Hall effect to the fractional quantum Hall effect~\cite{hall1}, quantum spin Hall effect~\cite{spin1,spin2}, quantum valley Hall effect~\cite{valley1}, and then extended many strange properties, such as bulk-boundary correspondence~\cite{bulk-boundary1,bulk-boundary2,bulk-boundary3,bulk-boundary4}, non-Hermitian skin effect~\cite{non-Hermitian1,non-Hermitian2,non-Hermitian3,non-Hermitian4,non-Hermitian5}, Floquet topology~\cite{Floquet1,Floquet2,Floquet3,Floquet4}, Anderson local~\cite{Anderson1,Anderson2}, Majorana fermion~\cite{Majorana1}, non-reciprocity~\cite{non-reciprocity1,non-reciprocity2}, non-Abelian topology~\cite{non-Abelian1,non-Abelian2,non-Abelian3}, hyperbolic metasurface topology~\cite{hyperbolic1}, et al. The combination of the excellent characteristics of the topological insulator of the post-electronic system and a variety of classical systems has pushed the research of classical systems to a whole new height, such as photons, phonons, circuits, cold atoms, mechanical and other systems~\cite{PPCCM1,PPCCM2,PPCCM3,PPCCM4,PPCCM5,PPCCM6,PPCCM7,PPCCM8}.
	
	The higher-order topological phase breaks through the bulk-boundary correspondence in traditional topology, and introduces a more abstract and complex concept involving higher-dimensional topological systems. Formally, the topological state of the lowest dimension contained in an $m$-D topological insulator system is $n$-D, so this topological insulator is called a $m$-$n$-order topological insulator.
	
	For example, if the lowest dimension of a $2$D system containing topological states is $0$D, then it is called $(2-0) = 2$-order topological insulator. Although the topological phase of matter originates from condensed matter systems, it has proven to be a ubiquitous property in a wide range of wave systems, and naturally, topological concepts can be extended to optical boson systems. As a classical optical analogue of electrons in quantum systems, photonic crystal platforms have the advantages of flexibility and diversity, making them ideal for realizing high-order topological insulators and exploring the basic mechanisms.~\cite{HO1,HO2,HO3,HO4,HO5,HO6}.
	
	The coupling size between split-rings can be conveniently controlled by adjusting the opening direction~\cite{SR1,SR2,SR3,SR4,SR5,SR6,SR7,SR8,SR9,SR10}. In photonic crystals that can be manipulated flexibly, this property can provide different rotational degrees of freedom for the exploration of photon platform topology. The research content of this paper is based on the rotation degrees of freedom of the four split-rings. Firstly, the emergence of photonic band gaps caused by the rotation of the split-rings is introduced. Then, the topological edge states are analyzed intuitively by constructing the interfaces of topologically trivial and non-trivial structures characterized by the bulk polarization. Then the higher-order corner states and its characteristic topological invariant - corner charge, are analyzed, and a brief summary is made.

	\section{MODEL}
	
	In this section, we construct the SSPC utilizing the principles of the classical $2$D Su-Schrieffer-Heeger (SSH) model from condensed matter physics.
	
\begin{figure*}[htbp]
		\centering
		\includegraphics[width=6in]{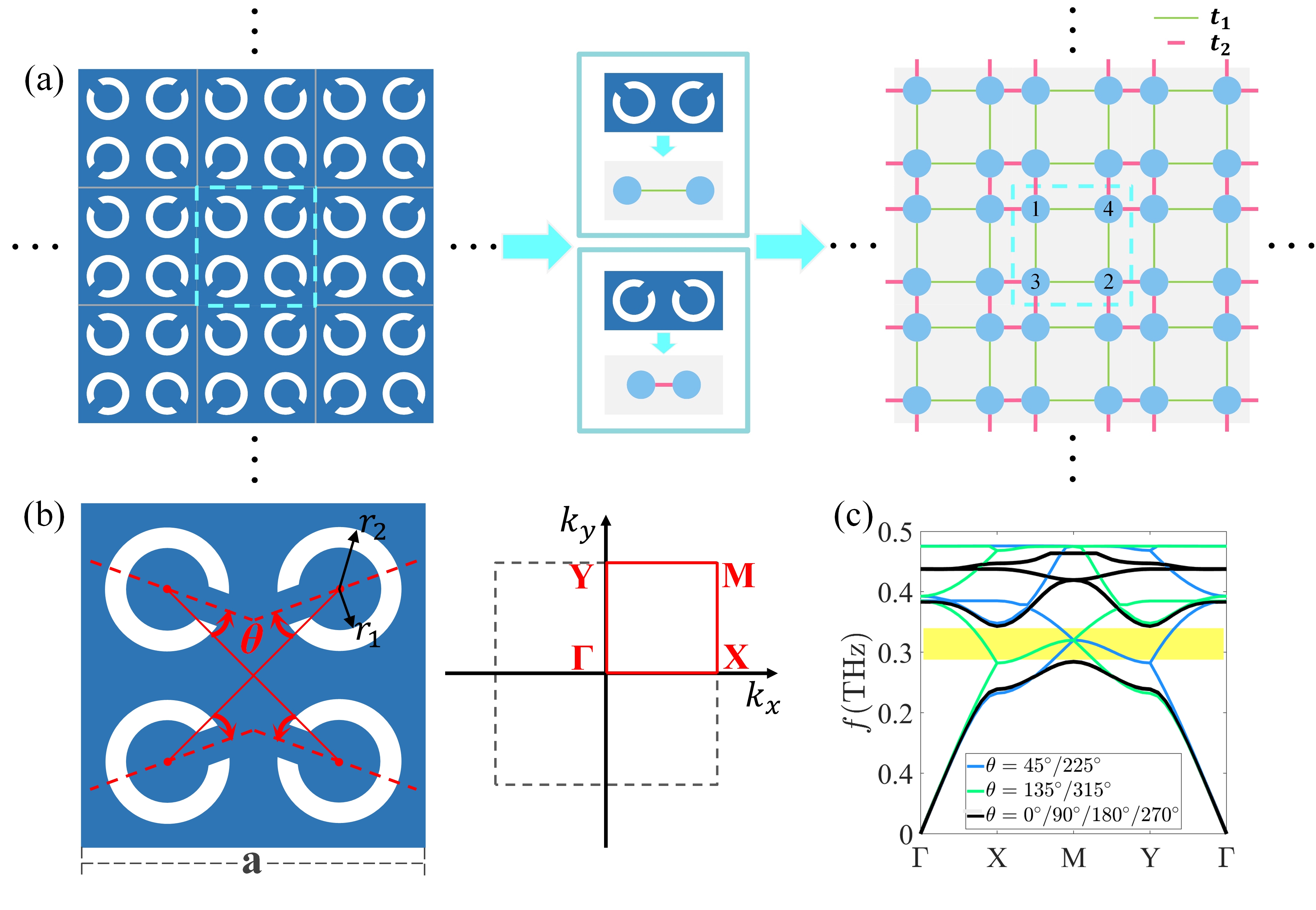}
		\caption{(a) Analogy between the SSPC and the $2$D SSH model in condensed matter. The opening direction between each two split-rings determines the coupling strength of the two rings. The coupling strength is smaller when the openings are oriented towards opposite directions compared to when they are oriented towards the same direction. Correspondingly, in the $2$D SSH model, the coupling strength $\omega$ of intracellular lattice points is smaller than $\upsilon$ of intercellular lattice points. (b) Unit cell and the FBZ of SSPC. The lattice constant is $a = 103 um$, the inner and outer radius of the split-ring $r_{1} = 0.13a$, $r_{2} = 0.18a$, the rotation angle is $\theta$, and the rotation directions and rotation centers are denoted by red solid arrows and red dots, respectively. The FBZ is marked with a black dotted line and the red solid rectangular box is the path to calculate the band structure. (c) Photonic bands of the SSPC for $\theta = 45^{\circ}/225^{\circ}$ (blue solid curves), $\theta = 135^{\circ}/315^{\circ}$ (green solid curve). The photonic bands of the SSPC for $\theta = 0^{\circ}/90^{\circ}/180^{\circ}/270^{\circ}$ is marked by a black solid curve, and the band gap marked in yellow appears.}
		\label{fig1}
	\end{figure*}

	As shown in Fig.~\ref{fig1}(a), we use split-rings to analogy tight-binding lattice, and consider the case that the opening orientations of two split-rings are different. The coupling strength when the opening orientations are far away from the center of the two rings is higher than that when the opening orientations are near the center of the two rings.
	
	In this way, the SSPC can be used to compare with the lattice model of the classical $2$D SSH model. The two cases respectively correspond to the weak coupling strength $\omega$ and strong coupling strength $\upsilon$. We simultaneously adjust the opening orientation of four split-rings in a unit cell to achieve the coupling strength regulation between the split-rings.
	
	In Fig.~\ref{fig1}(b), the rotation angles of the four split-rings are $\theta$; the lattice constant and inner and outer radius of the ring of SSPC are $a$, $r_{1}$ and $r_{2}$, respectively; the boundary between the ring and the background material is an ideal electrical conductor, and the relative dielectric constant of the background material is 11.56. By calculating the $\Gamma-X-M-Y-\Gamma$ path in the first Brillouin zone (FBZ), the band structure at different $\theta$ is obtained. When $\theta = 45^{\circ}/225^{\circ}$, there is no band gap, and there is band degeneracy in the $M-Y$ segment; when $\theta = 135^{\circ}/315^{\circ}$, there is no band gap in the $X-M$ segment; when $\theta = 0^{\circ}/90^{\circ}/180^{\circ}/270^{\circ}$, there exist a band gap between the first and second bands. As shown in Figure.~\ref{fig1}(c).
	
	In addition, the model can be characterized by the Hamiltonian matrix of a tight-binding approximation model: 
	
	\begin{equation}
		\begin{pmatrix}0&0&\omega_1+\upsilon_1\cdot e^{-i\cdot k_y}&\omega_2+\upsilon_2\cdot e^{-i\cdot k_x}\\0&0&\omega_2+\upsilon_2\cdot e^{i\cdot k_x}&\omega_1+\upsilon_1\cdot e^{i\cdot k_y}\\\omega_1+\upsilon_1\cdot e^{i\cdot k_y}&\omega_2+\upsilon_2\cdot e^{-i\cdot k_x}&0&0\\\omega_2+\upsilon_2\cdot e^{i\cdot k_x}&\omega_1+\upsilon_1\cdot e^{-i\cdot k_y}&0&0
		\end{pmatrix}
		\label{eq1}
	\end{equation}

	Where, $\omega_{1}=t_{0}+t_{1}\cdot\cos\left(\pi/4+\theta\right)$ represents the intracellular transition between grid points $1$ and $3$, $2$ and $4$, $\omega_{2}=t_{0}+t_{1}\cdot\cos\left(-\pi/4+\theta\right)$ represents the intracellular transition between grid points $1$ and $4$, $2$ and $3$, $\upsilon_{1}=t_{0}+t_{1}\cdot\cos\left(-3\pi/4+\theta\right)$ represents the intercellular transition between grid points $1$ and $3$, $2$ and $4$, $\upsilon_{2}=t_{0}+t_{1}\cdot\cos\left(3\pi/4+\theta\right)$ represents the intercellular transition between grid points $1$ and $4$, $2$ and $3$, where $t_{0} = 1$, $t_{1} = 0.8$, where the exponential silver $\theta$ is the periodic change factor of the modulation coupling intensity. The period is $[0,2\pi]$, the band structure described by the Hamiltonian can also open a band gap between the first and second bands (Appendix 1).

	\section{Topological invariant}
	
	The topological properties of the different shapes of the preceding band structure at different rotation angles can be described by the parities at the highly symmetric point $X/Y$.~\cite{SR2}

	Where, $p_{x/y}$ represents the volume polarization in the $x/y$ direction, and $\eta_m(\Gamma/X/Y)$ represents the parity of the characteristic state at the highly symmetric point $\Gamma/X/Y$ of the m band. The parity is $+1$ and the parity is $-1$. The band gap of the model in this paper occurs between the first and second energy bands, so only the case of $m=1$ is considered here
	
\begin{figure}[htbp]
			\centering
			\includegraphics[width=\columnwidth]{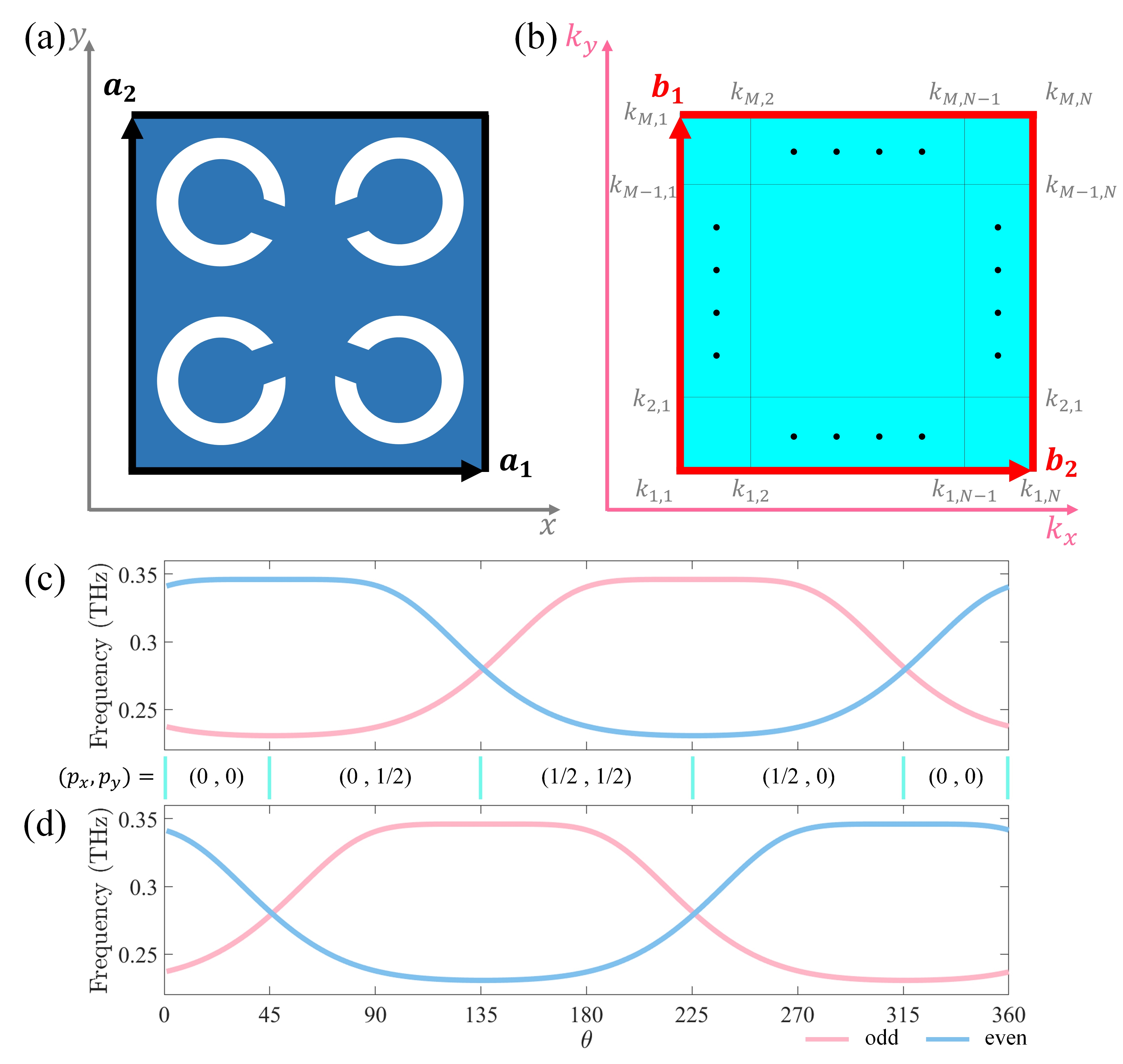}
			\caption{(a) The unit cell is marked with a solid black line, and the real-space lattice vector are marked by black arrows. (b) The FBZ is marked by a solid red line and the reciprocal lattice vectors are marked by red arrows. The FBZ is divided into $(M-1)*(N-1)$ regions. Phase transition diagram of the bottom two eigenvalues as a function of $\theta$ (c) at $X$ and (d) at $Y$. Light red and light blue curves respectively represent the odd parity and even parity of the eigenstates. By analyzing the parity of the first eigenvalue, the corresponding bulk polarization ($p_{x}$, $p_{y}$) can be obtained. The interval of 1-360° consists of four sections with different bulk polarization, which are: $\theta=0^{\circ}-45^{\circ}$ and $315^{\circ}-360^{\circ}$ is $(0,0)$, $45^\theta={\circ}-135^{\circ}$ is $(0,1/2)$, $\theta=135^{\circ}-225^{\circ}$ is $(1/2,1/2)$, $\theta=225^{\circ}-315^{\circ}$ is $(1/2,0)$.}
			\label{fig2}
	\end{figure}

	As shown in Fig.~\ref{fig2}(c) and d), the two figures are phase transition diagrams of the first and second frequencies at highly symmetric points $X$ and $Y$ with respect to the Angle of $\theta$. In a period of change of $\theta$ ($0^{\circ}-360^{\circ}$), the first frequency and the second frequency experience two degeneracy breaks, and their parity also changes with $\theta$. This results in four intervals with different topological properties within a period of change.
	
	$\eta(\Gamma) = +1$ over a period of change. In the interval $0^{\circ}-45^{\circ}$ and $315^{\circ}-360^{\circ}$ $\eta(X) = \eta(Y) = +1$, in the interval $45^{\circ}-135^{\circ}$ $\eta(X) = +1$, $\eta(Y) = -1$, in the interval $135^{\circ}-225^{\circ}$ $\eta(X) = \eta(Y) = -1$, and in the interval $225^{\circ}-315^{\circ}$ $\eta(X) = -1$, $\eta(Y) = +1$, it is easy to know that the bulk polarization of these four regions is $\left(p_{x},p_{y}\right)=(0,0)\mid(0,\frac{1}{2})\mid(\frac{1}{2},\frac{1}{2})\mid(\frac{1}{2},0)$ respectively. Next, we will use numerical calculation of topological invariant -- volume polarization to verify the results of the above analysis
	
	In this part, we will analyze the related topological invariants of SSPC, namely bulk polarization derived from Wilson loop and edge polarization derive from corner charge.
	
	Based on the Berry phase of electrons in condensed matter, the topological polarization theory of Bloch photons in photonic crystals has been developed. Here, since the SSPC satisfies mirror symmetry, the Wannier center of the states can be shifted by $0$ or $1/2$ lattice constant from the center of the unit cell by designing the structure parameters, while these two deviations mark the trivial and nontrivial polarization, respectively. In addition, the finite polarization of the bulk states results in the emergence of the edge states in the band gap, while the finite polarization of the edge states results in the emergence of the corner states~\cite{HO1,HO3,HO4,HO5}. 
	
	Firstly, we define the real space lattice basis vector of SSPC: $a_{1} = (a, 0, 0)$ and $a_{2} = (0, a, 0)$. 
	To consider $a$ unit vector $a_{3} = (0, 0, 1)$ in the z direction and solve for the reciprocal lattice vector: $b_1=\frac{2\pi\big(a_2\times a_3\big)}{a_1\cdot\big(a_2\times a_3\big)}=\bigg(\frac{2\pi}a,0\bigg)$, $\quad b_2=\frac{2\pi\big(a_3\times a_1\big)}{a_1\cdot\big(a_2\times a_3\big)}=\bigg(0,\frac{2\pi}a\bigg)$.

	The corresponding FBZ can be obtained as shown in Fig.~\ref{fig2}(b). The FBZ is divided into $M\dot N$ $k$ points along $b_{1}$ and  $b_{2}$ directions respectively (we use $M = N$ in all calculations). Then, the Wilson loop can be defined:
	
	\begin{equation}
		W_{\omega,\upsilon}^{x,y}=F_{\omega,\upsilon}^{x,y}\left(k\right)\cdot F_{\omega,\upsilon}^{x,y}\left(k+\Delta k\right)\cdots F_{\omega,\upsilon}^{x,y}\left(k+\left(N-1\right)\cdot\Delta k\right)
		\label{eq2}
	\end{equation}

	where $\left.F_{\omega,\upsilon}^{x,y}\left(k\right)=\left\langle\varphi_{\omega}^{x,y}\left(k\right)\right|\varphi_{\upsilon}^{x,y}\left(k+\Delta k\right)\right\rangle $ are the Wilson loop elements, in which $\Delta k=\frac{2\pi}{N}$, the $\left|\varphi_{\omega,\upsilon}^{x,y}(k+\Delta k)\right\rangle $ is the periodic part of the wave functions of the $\omega$-th, $\upsilon$-th order band with wave vector $k$ along $x$ and $y$ direction, respectively. $\varepsilon\left(r\right)$ is the position-dependent electric permittivity.

	The Wilson loop operator along   and   directions can be diagonalized as: $\left.\left.W_{\omega,\upsilon}^y\left|\chi\left(k_x\right)\right\rangle=\exp\left(i\cdot2\pi\Re_y\left(k_x\right)\right)\right|\chi\left(k_x\right)\right\rangle $ and $\left.W_{\omega,\nu}^{x}\left|\left.\chi\left(k_{y}\right)\right)=\exp\left(i\cdot2\pi\Re_{x}\left(k_{y}\right)\right)\right|\chi\left(k_{y}\right)\right\rangle $, where $\left|\chi(k_x)\right\rangle$ and $\left|\chi(k_y)\right\rangle $ are the eigenvectors which depends on the Wilson loop, and the phase $\Re_{y}(k_{x})$ and $\Re_{x}(k_{y})$ are the elements that form the Wilson loop.
	
	Now, the bulk polarization can be defined as
	
	\begin{equation}
	\begin{aligned}p_x&=\frac{1}{N}\sum_{k_x}\Re_y\left(k_x\right)\\p_y&=\frac{1}{N}\sum_{k_y}\Re_x\left(k_y\right)\end{aligned}
	\label{eq3}
	\end{equation}
	
	We calculated the four band gap regions mentioned above by using the method of calculation   and   above. The results obtained are consistent with the results of parity analysis above, and the polarization of the four band gap regions is $(p_{x},p_{y})=(0,0)|\left(0, 1/2\right)|(1/2,  1/2)|(1/2, 0)$.

	\section{Topological edge and corner states}
	
\begin{figure}[htbp]
		\centering
		\includegraphics[width=\columnwidth]{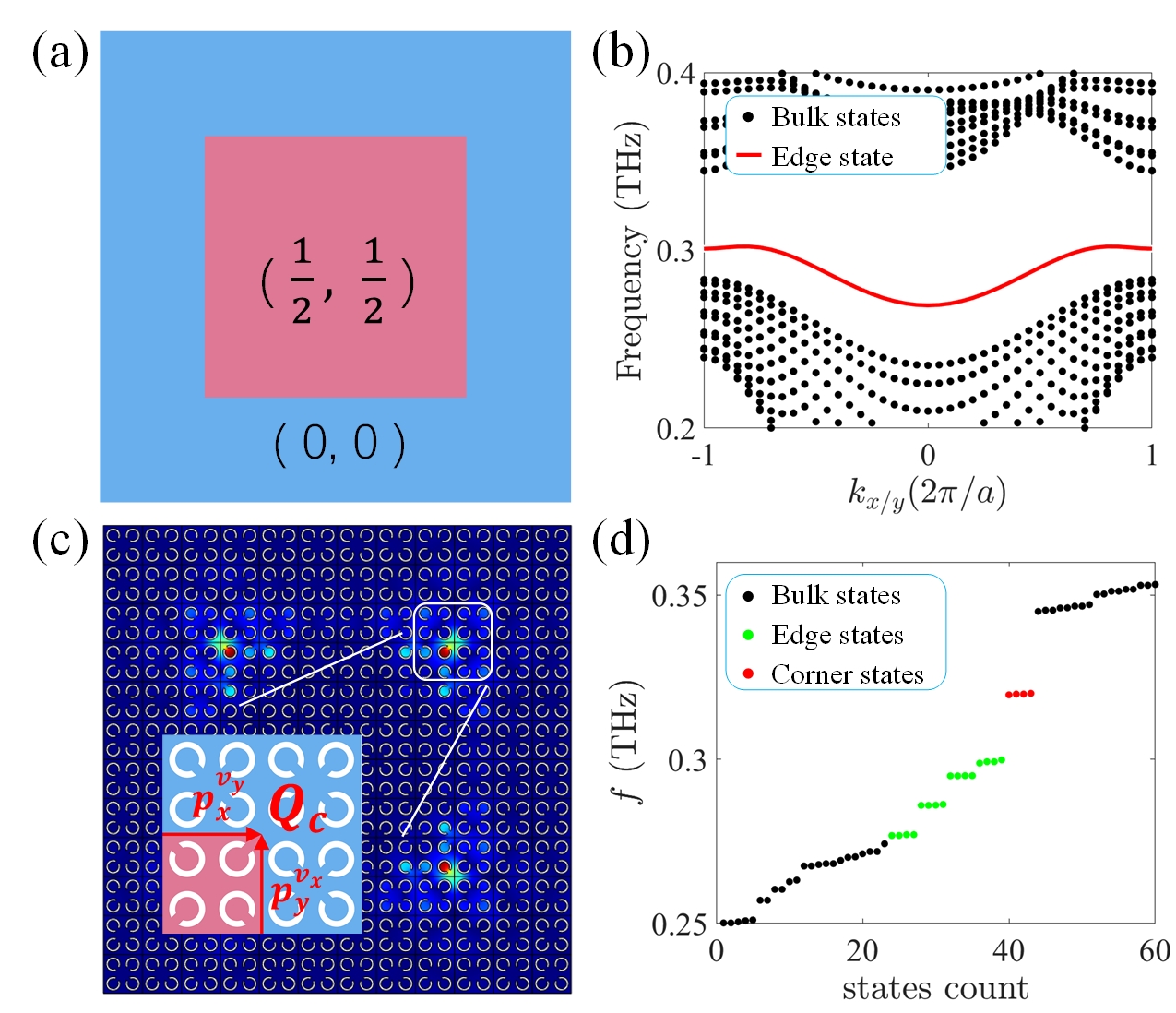}
		\caption{(a) Schematic diagram of the SSPC finite model. Light red and light blue represent the SSPC region of the non-trivial topology and the trivial topology respectively, and the corresponding bulk polarization are $(p_x,p_y)=(0,0)$ and $(p_x,p_y)=(1/2,1/2)$. (b) Photonic bands of the supercell which composed of $(p_x,p_y)=(0,0)$ and $(p_x,p_y)=(1/2,1/2)$. The black points are the bulk states, and the red solid curve is the edge state. (c) $|TE|$ field distribution of corner state, represents the corner charge interpretation of the corner state: the polarization of the edge leads to the appearance of the topological corner state. (d) The eigenfrequency diagram corresponding to the finite model. The red dots, green dots and black dots represent corner states, edge states and bulk states, respectively.}
		\label{fig3}
	\end{figure}

	Our proposed model exhibits second-order corner states. Here, we construct a finite lattice system using models with $(p_x,p_y)=(0,0)$ and $(p_x,p_y)=(1/2,1/2)$, as shown in Fig.~\ref{fig3}(a). We surround the model with $(p_x,p_y)=(1/2,1/2)$ using a structure with $(p_x,p_y)=(0,0)$. The internal structure has a bulk polarization of $1/2$ in both the $x$ and $y$ directions, which represents a nontrivial topological phase. At the boundary interface with the trivial topological structure, the existence of topological edge states can be supported by the non-zero bulk polarization. The band structures of the horizontal and vertical supercells, shown in Fig.~\ref{fig3}(b), demonstrate the existence of edge states within the band gap. We then calculate the characteristic frequencies of this finite system, as shown in Fig.~\ref{fig3}(d). Besides the topological edge states, there are four topological corner states within the band gap. The field distribution of the corner states is shown in Fig.~\ref{fig3}(c), which is localized in the four corners of the model with $(p_x,p_y)=(1/2,1/2)$. If the topological edge states can be understood as "edge of the bulk," then the topological properties of the corner states can be understood as "edge of the edge." Due to the bulk polarizations of $1/2$ in both $x$ and $y$ directions of the enclosed lattice model, there can be a description of the corner charge $Q_{c}$ of the corner mode:
	
	\begin{equation}
		Q_{c}=4\cdot P_{x}\cdot P_{y}
		\label{eq4}
	\end{equation}
	
	When the corner charge of the wrapped model is non-zero, there will be a topological angular mode, that is to say, only when the bulk polarization of wrapped model is $(p_x,p_y)=(1/2,1/2)$, $Q_{c} = 1$, the topological corner states will appear.

\begin{figure*}[htbp]
		\centering
		\includegraphics[width=5.2in]{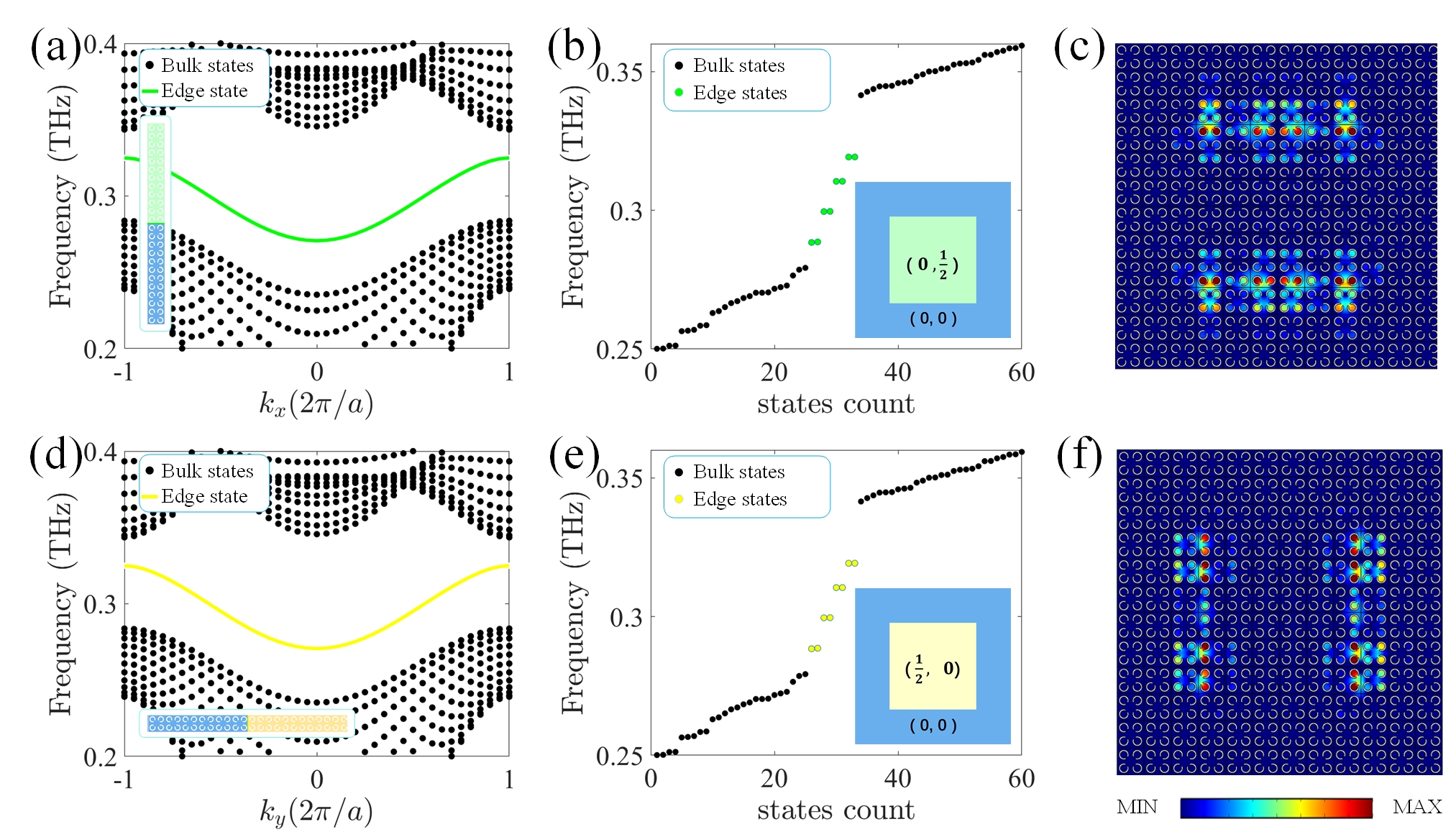}
		\caption{(a) Schematic diagram of the vertical supercell and its band structure (composed of $(p_x,p_y)=(0,0)$ (light blue area) and $(p_x,p_y)=(0,1/2)$ (light green area)). The black dots represent the bulk states and the solid green curve represents the edge state. (b) Eigenfrequencies corresponding to the finite models. Black dots and green dots represent bulk and edge states. (c) $|TE|$ field distribution diagram of edge states. (d) The horizontal supercell diagram and its band structure (composed of $(p_x,p_y)=(0,0)$ and $(p_x,p_y)=(1/2,0)$ (light yellow area)), with black dots representing bulk states and green solid curve representing edge state. (e) Eigenfrequencies corresponding to the finite models. Black dots and yellow dots represent bulk and edge states. (f) $|TE|$ field distribution diagram of edge states.}
		\label{fig4}
	\end{figure*}

	When $(p_x,p_y)=(0,0)$ and $(p_x,p_y)=(1/2,1/2)$, the topology trivial and non-trivial properties of the model are corresponding respectively. Such topological distinction can be corresponding to $p_x$ and $p_y$ and they will not affect each other. In other words, when the two models with $(p_x,p_y)=(0,0)$ and $(p_x,p_y)=(0,1/2)$ are used to form the interface system, since the $x$ direction is only topologically trivial, while the y direction is topologically trivial and non-trivial, there does not exist topological edge state in the $x$ direction. As shown in Fig.~\ref{fig4}(a-c), in the $y$ direction, there exist topological edge states described by non-zero bulk polarization. Similarly, in the interface system consisting of two models with $(p_x, p_y) = (0,0)$ and $(p_x, p_y) = (1/2, 0)$, there only exist topological edge states described by non-zero bulk polarization in the $x$ direction, but does not in the y direction, as shown in Fig.~\ref{fig4}(d-f).
	
	\section{Conclusion and discussions}
	We propose a photonic crystal model composed of four split-rings. The coupling strength between the split-rings can be corresponded by the lattice point transition in condensed matter. When the four split-rings are rotated at the same time, there exist two "open-degenerate-open" processes between the first frequency and the second frequency at the highly symmetric points $X$ and $Y$ in a 2$\pi$ rotation period. During the period, there exist four types band gaps with different topological properties, corresponding to different bulk polarization. When the bulk polarization in the $x/y$ direction are equal to $1/2$, topological edge state appearing at the $x/y$ interface composed of topologically trivial structures. In addition, when the bulk polarization in the $x$ and $y$ directions are both equal to $1/2$, "edge states of edge states" appear, that is, corner states. Our model can be further extended to other classical systems, and the concept of "coupling can be changed by rotation" can be applied to dynamic continuously modulated systems, such as Floquet topology and synthetic dimension. Our research content enriches the research content of photonic crystal higher-order topology and broadens the road for the follow-up research.
	
	\section*{ACKNOWLEDGMENTS}
	This work was supported by the National Natural Science Foundation of China (Grant numbers 61865009, 61927813).
	
	\bibliography{main.bib}

\begin{thebibliography}{50}%
\makeatletter
\providecommand \@ifxundefined [1]{%
 \@ifx{#1\undefined}
}%
\providecommand \@ifnum [1]{%
 \ifnum #1\expandafter \@firstoftwo
 \else \expandafter \@secondoftwo
 \fi
}%
\providecommand \@ifx [1]{%
 \ifx #1\expandafter \@firstoftwo
 \else \expandafter \@secondoftwo
 \fi
}%
\providecommand \natexlab [1]{#1}%
\providecommand \enquote  [1]{``#1''}%
\providecommand \bibnamefont  [1]{#1}%
\providecommand \bibfnamefont [1]{#1}%
\providecommand \citenamefont [1]{#1}%
\providecommand \href@noop [0]{\@secondoftwo}%
\providecommand \href [0]{\begingroup \@sanitize@url \@href}%
\providecommand \@href[1]{\@@startlink{#1}\@@href}%
\providecommand \@@href[1]{\endgroup#1\@@endlink}%
\providecommand \@sanitize@url [0]{\catcode `\\12\catcode `\$12\catcode
  `\&12\catcode `\#12\catcode `\^12\catcode `\_12\catcode `\%12\relax}%
\providecommand \@@startlink[1]{}%
\providecommand \@@endlink[0]{}%
\providecommand \url  [0]{\begingroup\@sanitize@url \@url }%
\providecommand \@url [1]{\endgroup\@href {#1}{\urlprefix }}%
\providecommand \urlprefix  [0]{URL }%
\providecommand \Eprint [0]{\href }%
\providecommand \doibase [0]{https://doi.org/}%
\providecommand \selectlanguage [0]{\@gobble}%
\providecommand \bibinfo  [0]{\@secondoftwo}%
\providecommand \bibfield  [0]{\@secondoftwo}%
\providecommand \translation [1]{[#1]}%
\providecommand \BibitemOpen [0]{}%
\providecommand \bibitemStop [0]{}%
\providecommand \bibitemNoStop [0]{.\EOS\space}%
\providecommand \EOS [0]{\spacefactor3000\relax}%
\providecommand \BibitemShut  [1]{\csname bibitem#1\endcsname}%
\let\auto@bib@innerbib\@empty
\bibitem [{\citenamefont {Klitzing}\ \emph {et~al.}(1980)\citenamefont
  {Klitzing}, \citenamefont {Dorda},\ and\ \citenamefont {Pepper}}]{hall1}%
  \BibitemOpen
  \bibfield  {author} {\bibinfo {author} {\bibfnamefont {K.~v.}\ \bibnamefont
  {Klitzing}}, \bibinfo {author} {\bibfnamefont {G.}~\bibnamefont {Dorda}},\
  and\ \bibinfo {author} {\bibfnamefont {M.}~\bibnamefont {Pepper}},\
  }\bibfield  {title} {\bibinfo {title} {New method for high-accuracy
  determination of the fine-structure constant based on quantized hall
  resistance},\ }\href {https://doi.org/10.1103/PhysRevLett.45.494} {\bibfield
  {journal} {\bibinfo  {journal} {Phys. Rev. Lett.}\ }\textbf {\bibinfo
  {volume} {45}},\ \bibinfo {pages} {494} (\bibinfo {year} {1980})}\BibitemShut
  {NoStop}%
\bibitem [{\citenamefont {Kane}\ and\ \citenamefont
  {Mele}(2005{\natexlab{a}})}]{spin1}%
  \BibitemOpen
  \bibfield  {author} {\bibinfo {author} {\bibfnamefont {C.~L.}\ \bibnamefont
  {Kane}}\ and\ \bibinfo {author} {\bibfnamefont {E.~J.}\ \bibnamefont
  {Mele}},\ }\bibfield  {title} {\bibinfo {title} {Quantum spin hall effect in
  graphene},\ }\href {https://doi.org/10.1103/PhysRevLett.95.226801} {\bibfield
   {journal} {\bibinfo  {journal} {Phys. Rev. Lett.}\ }\textbf {\bibinfo
  {volume} {95}},\ \bibinfo {pages} {226801} (\bibinfo {year}
  {2005}{\natexlab{a}})}\BibitemShut {NoStop}%
\bibitem [{\citenamefont {Kane}\ and\ \citenamefont
  {Mele}(2005{\natexlab{b}})}]{spin2}%
  \BibitemOpen
  \bibfield  {author} {\bibinfo {author} {\bibfnamefont {C.~L.}\ \bibnamefont
  {Kane}}\ and\ \bibinfo {author} {\bibfnamefont {E.~J.}\ \bibnamefont
  {Mele}},\ }\bibfield  {title} {\bibinfo {title} {${Z}_{2}$ topological order
  and the quantum spin hall effect},\ }\href
  {https://doi.org/10.1103/PhysRevLett.95.146802} {\bibfield  {journal}
  {\bibinfo  {journal} {Phys. Rev. Lett.}\ }\textbf {\bibinfo {volume} {95}},\
  \bibinfo {pages} {146802} (\bibinfo {year} {2005}{\natexlab{b}})}\BibitemShut
  {NoStop}%
\bibitem [{\citenamefont {Xiao}\ \emph {et~al.}(2007)\citenamefont {Xiao},
  \citenamefont {Yao},\ and\ \citenamefont {Niu}}]{valley1}%
  \BibitemOpen
  \bibfield  {author} {\bibinfo {author} {\bibfnamefont {D.}~\bibnamefont
  {Xiao}}, \bibinfo {author} {\bibfnamefont {W.}~\bibnamefont {Yao}},\ and\
  \bibinfo {author} {\bibfnamefont {Q.}~\bibnamefont {Niu}},\ }\bibfield
  {title} {\bibinfo {title} {Valley-contrasting physics in graphene: Magnetic
  moment and topological transport},\ }\href
  {https://doi.org/10.1103/PhysRevLett.99.236809} {\bibfield  {journal}
  {\bibinfo  {journal} {Phys. Rev. Lett.}\ }\textbf {\bibinfo {volume} {99}},\
  \bibinfo {pages} {236809} (\bibinfo {year} {2007})}\BibitemShut {NoStop}%
\bibitem [{\citenamefont {Chen}\ \emph {et~al.}(2023)\citenamefont {Chen},
  \citenamefont {Liu}, \citenamefont {Wu}, \citenamefont {Su}, \citenamefont
  {Ding}, \citenamefont {Qin}, \citenamefont {Wang}, \citenamefont {Zhang},
  \citenamefont {He}, \citenamefont {Wang}, \citenamefont {Lu}, \citenamefont
  {Li}, \citenamefont {Sanders}, \citenamefont {Liu},\ and\ \citenamefont
  {Pan}}]{bulk-boundary1}%
  \BibitemOpen
  \bibfield  {author} {\bibinfo {author} {\bibfnamefont {C.}~\bibnamefont
  {Chen}}, \bibinfo {author} {\bibfnamefont {R.-Z.}\ \bibnamefont {Liu}},
  \bibinfo {author} {\bibfnamefont {J.}~\bibnamefont {Wu}}, \bibinfo {author}
  {\bibfnamefont {Z.-E.}\ \bibnamefont {Su}}, \bibinfo {author} {\bibfnamefont
  {X.}~\bibnamefont {Ding}}, \bibinfo {author} {\bibfnamefont {J.}~\bibnamefont
  {Qin}}, \bibinfo {author} {\bibfnamefont {L.}~\bibnamefont {Wang}}, \bibinfo
  {author} {\bibfnamefont {W.-W.}\ \bibnamefont {Zhang}}, \bibinfo {author}
  {\bibfnamefont {Y.}~\bibnamefont {He}}, \bibinfo {author} {\bibfnamefont
  {X.-L.}\ \bibnamefont {Wang}}, \bibinfo {author} {\bibfnamefont {C.-Y.}\
  \bibnamefont {Lu}}, \bibinfo {author} {\bibfnamefont {L.}~\bibnamefont {Li}},
  \bibinfo {author} {\bibfnamefont {B.~C.}\ \bibnamefont {Sanders}}, \bibinfo
  {author} {\bibfnamefont {X.-J.}\ \bibnamefont {Liu}},\ and\ \bibinfo {author}
  {\bibfnamefont {J.-W.}\ \bibnamefont {Pan}},\ }\bibfield  {title} {\bibinfo
  {title} {Berry curvature and bulk-boundary correspondence from transport
  measurement for photonic chern bands},\ }\href
  {https://doi.org/10.1103/PhysRevLett.131.133601} {\bibfield  {journal}
  {\bibinfo  {journal} {Phys. Rev. Lett.}\ }\textbf {\bibinfo {volume} {131}},\
  \bibinfo {pages} {133601} (\bibinfo {year} {2023})}\BibitemShut {NoStop}%
\bibitem [{\citenamefont {Edvardsson}\ \emph {et~al.}(2019)\citenamefont
  {Edvardsson}, \citenamefont {Kunst},\ and\ \citenamefont
  {Bergholtz}}]{bulk-boundary2}%
  \BibitemOpen
  \bibfield  {author} {\bibinfo {author} {\bibfnamefont {E.}~\bibnamefont
  {Edvardsson}}, \bibinfo {author} {\bibfnamefont {F.~K.}\ \bibnamefont
  {Kunst}},\ and\ \bibinfo {author} {\bibfnamefont {E.~J.}\ \bibnamefont
  {Bergholtz}},\ }\bibfield  {title} {\bibinfo {title} {Non-hermitian
  extensions of higher-order topological phases and their biorthogonal
  bulk-boundary correspondence},\ }\href
  {https://doi.org/10.1103/PhysRevB.99.081302} {\bibfield  {journal} {\bibinfo
  {journal} {Phys. Rev. B}\ }\textbf {\bibinfo {volume} {99}},\ \bibinfo
  {pages} {081302} (\bibinfo {year} {2019})}\BibitemShut {NoStop}%
\bibitem [{\citenamefont {Mas\l{}owski}\ and\ \citenamefont
  {Sedlmayr}(2023)}]{bulk-boundary3}%
  \BibitemOpen
  \bibfield  {author} {\bibinfo {author} {\bibfnamefont {T.}~\bibnamefont
  {Mas\l{}owski}}\ and\ \bibinfo {author} {\bibfnamefont {N.}~\bibnamefont
  {Sedlmayr}},\ }\bibfield  {title} {\bibinfo {title} {Dynamical bulk-boundary
  correspondence and dynamical quantum phase transitions in higher-order
  topological insulators},\ }\href
  {https://doi.org/10.1103/PhysRevB.108.094306} {\bibfield  {journal} {\bibinfo
   {journal} {Phys. Rev. B}\ }\textbf {\bibinfo {volume} {108}},\ \bibinfo
  {pages} {094306} (\bibinfo {year} {2023})}\BibitemShut {NoStop}%
\bibitem [{\citenamefont {Xiao}\ \emph {et~al.}(2020)\citenamefont {Xiao},
  \citenamefont {Deng}, \citenamefont {Wang}, \citenamefont {Zhu},
  \citenamefont {Wang}, \citenamefont {Yi},\ and\ \citenamefont
  {Xue}}]{bulk-boundary4}%
  \BibitemOpen
  \bibfield  {author} {\bibinfo {author} {\bibfnamefont {L.}~\bibnamefont
  {Xiao}}, \bibinfo {author} {\bibfnamefont {T.}~\bibnamefont {Deng}}, \bibinfo
  {author} {\bibfnamefont {K.}~\bibnamefont {Wang}}, \bibinfo {author}
  {\bibfnamefont {G.}~\bibnamefont {Zhu}}, \bibinfo {author} {\bibfnamefont
  {Z.}~\bibnamefont {Wang}}, \bibinfo {author} {\bibfnamefont {W.}~\bibnamefont
  {Yi}},\ and\ \bibinfo {author} {\bibfnamefont {P.}~\bibnamefont {Xue}},\
  }\bibfield  {title} {\bibinfo {title} {Non-hermitian bulk--boundary
  correspondence in quantum dynamics},\ }\href@noop {} {\bibfield  {journal}
  {\bibinfo  {journal} {Nature Physics}\ }\textbf {\bibinfo {volume} {16}},\
  \bibinfo {pages} {761} (\bibinfo {year} {2020})}\BibitemShut {NoStop}%
\bibitem [{\citenamefont {Yao}\ and\ \citenamefont
  {Wang}(2018)}]{non-Hermitian1}%
  \BibitemOpen
  \bibfield  {author} {\bibinfo {author} {\bibfnamefont {S.}~\bibnamefont
  {Yao}}\ and\ \bibinfo {author} {\bibfnamefont {Z.}~\bibnamefont {Wang}},\
  }\bibfield  {title} {\bibinfo {title} {Edge states and topological invariants
  of non-hermitian systems},\ }\href
  {https://doi.org/10.1103/PhysRevLett.121.086803} {\bibfield  {journal}
  {\bibinfo  {journal} {Phys. Rev. Lett.}\ }\textbf {\bibinfo {volume} {121}},\
  \bibinfo {pages} {086803} (\bibinfo {year} {2018})}\BibitemShut {NoStop}%
\bibitem [{\citenamefont {Li}\ \emph {et~al.}(2022)\citenamefont {Li},
  \citenamefont {Liang}, \citenamefont {Wang}, \citenamefont {Lu},\ and\
  \citenamefont {Liu}}]{non-Hermitian2}%
  \BibitemOpen
  \bibfield  {author} {\bibinfo {author} {\bibfnamefont {Y.}~\bibnamefont
  {Li}}, \bibinfo {author} {\bibfnamefont {C.}~\bibnamefont {Liang}}, \bibinfo
  {author} {\bibfnamefont {C.}~\bibnamefont {Wang}}, \bibinfo {author}
  {\bibfnamefont {C.}~\bibnamefont {Lu}},\ and\ \bibinfo {author}
  {\bibfnamefont {Y.-C.}\ \bibnamefont {Liu}},\ }\bibfield  {title} {\bibinfo
  {title} {Gain-loss-induced hybrid skin-topological effect},\ }\href
  {https://doi.org/10.1103/PhysRevLett.128.223903} {\bibfield  {journal}
  {\bibinfo  {journal} {Phys. Rev. Lett.}\ }\textbf {\bibinfo {volume} {128}},\
  \bibinfo {pages} {223903} (\bibinfo {year} {2022})}\BibitemShut {NoStop}%
\bibitem [{\citenamefont {Zhang}\ \emph {et~al.}(2022)\citenamefont {Zhang},
  \citenamefont {Yang},\ and\ \citenamefont {Fang}}]{non-Hermitian3}%
  \BibitemOpen
  \bibfield  {author} {\bibinfo {author} {\bibfnamefont {K.}~\bibnamefont
  {Zhang}}, \bibinfo {author} {\bibfnamefont {Z.}~\bibnamefont {Yang}},\ and\
  \bibinfo {author} {\bibfnamefont {C.}~\bibnamefont {Fang}},\ }\bibfield
  {title} {\bibinfo {title} {Universal non-hermitian skin effect in two and
  higher dimensions},\ }\href@noop {} {\bibfield  {journal} {\bibinfo
  {journal} {Nature communications}\ }\textbf {\bibinfo {volume} {13}},\
  \bibinfo {pages} {2496} (\bibinfo {year} {2022})}\BibitemShut {NoStop}%
\bibitem [{\citenamefont {Zhang}\ \emph {et~al.}(2021)\citenamefont {Zhang},
  \citenamefont {Tian}, \citenamefont {Jiang}, \citenamefont {Lu},\ and\
  \citenamefont {Chen}}]{non-Hermitian4}%
  \BibitemOpen
  \bibfield  {author} {\bibinfo {author} {\bibfnamefont {X.}~\bibnamefont
  {Zhang}}, \bibinfo {author} {\bibfnamefont {Y.}~\bibnamefont {Tian}},
  \bibinfo {author} {\bibfnamefont {J.-H.}\ \bibnamefont {Jiang}}, \bibinfo
  {author} {\bibfnamefont {M.-H.}\ \bibnamefont {Lu}},\ and\ \bibinfo {author}
  {\bibfnamefont {Y.-F.}\ \bibnamefont {Chen}},\ }\bibfield  {title} {\bibinfo
  {title} {Observation of higher-order non-hermitian skin effect},\ }\href@noop
  {} {\bibfield  {journal} {\bibinfo  {journal} {Nature communications}\
  }\textbf {\bibinfo {volume} {12}},\ \bibinfo {pages} {5377} (\bibinfo {year}
  {2021})}\BibitemShut {NoStop}%
\bibitem [{\citenamefont {Xin}\ \emph {et~al.}(2023)\citenamefont {Xin},
  \citenamefont {Song}, \citenamefont {Wu}, \citenamefont {Lin}, \citenamefont
  {Zhu},\ and\ \citenamefont {Li}}]{non-Hermitian5}%
  \BibitemOpen
  \bibfield  {author} {\bibinfo {author} {\bibfnamefont {H.}~\bibnamefont
  {Xin}}, \bibinfo {author} {\bibfnamefont {W.}~\bibnamefont {Song}}, \bibinfo
  {author} {\bibfnamefont {S.}~\bibnamefont {Wu}}, \bibinfo {author}
  {\bibfnamefont {Z.}~\bibnamefont {Lin}}, \bibinfo {author} {\bibfnamefont
  {S.}~\bibnamefont {Zhu}},\ and\ \bibinfo {author} {\bibfnamefont
  {T.}~\bibnamefont {Li}},\ }\bibfield  {title} {\bibinfo {title} {Manipulating
  the non-hermitian skin effect in optical ring resonators},\ }\href
  {https://doi.org/10.1103/PhysRevB.107.165401} {\bibfield  {journal} {\bibinfo
   {journal} {Phys. Rev. B}\ }\textbf {\bibinfo {volume} {107}},\ \bibinfo
  {pages} {165401} (\bibinfo {year} {2023})}\BibitemShut {NoStop}%
\bibitem [{\citenamefont {Rechtsman}\ \emph {et~al.}(2013)\citenamefont
  {Rechtsman}, \citenamefont {Zeuner}, \citenamefont {Plotnik}, \citenamefont
  {Lumer}, \citenamefont {Podolsky}, \citenamefont {Dreisow}, \citenamefont
  {Nolte}, \citenamefont {Segev},\ and\ \citenamefont {Szameit}}]{Floquet1}%
  \BibitemOpen
  \bibfield  {author} {\bibinfo {author} {\bibfnamefont {M.~C.}\ \bibnamefont
  {Rechtsman}}, \bibinfo {author} {\bibfnamefont {J.~M.}\ \bibnamefont
  {Zeuner}}, \bibinfo {author} {\bibfnamefont {Y.}~\bibnamefont {Plotnik}},
  \bibinfo {author} {\bibfnamefont {Y.}~\bibnamefont {Lumer}}, \bibinfo
  {author} {\bibfnamefont {D.}~\bibnamefont {Podolsky}}, \bibinfo {author}
  {\bibfnamefont {F.}~\bibnamefont {Dreisow}}, \bibinfo {author} {\bibfnamefont
  {S.}~\bibnamefont {Nolte}}, \bibinfo {author} {\bibfnamefont
  {M.}~\bibnamefont {Segev}},\ and\ \bibinfo {author} {\bibfnamefont
  {A.}~\bibnamefont {Szameit}},\ }\bibfield  {title} {\bibinfo {title}
  {Photonic floquet topological insulators},\ }\href@noop {} {\bibfield
  {journal} {\bibinfo  {journal} {Nature}\ }\textbf {\bibinfo {volume} {496}},\
  \bibinfo {pages} {196} (\bibinfo {year} {2013})}\BibitemShut {NoStop}%
\bibitem [{\citenamefont {Ke}\ \emph {et~al.}(2023)\citenamefont {Ke},
  \citenamefont {Wen}, \citenamefont {Zhao},\ and\ \citenamefont
  {Wang}}]{Floquet2}%
  \BibitemOpen
  \bibfield  {author} {\bibinfo {author} {\bibfnamefont {S.}~\bibnamefont
  {Ke}}, \bibinfo {author} {\bibfnamefont {W.}~\bibnamefont {Wen}}, \bibinfo
  {author} {\bibfnamefont {D.}~\bibnamefont {Zhao}},\ and\ \bibinfo {author}
  {\bibfnamefont {Y.}~\bibnamefont {Wang}},\ }\bibfield  {title} {\bibinfo
  {title} {Floquet engineering of the non-hermitian skin effect in photonic
  waveguide arrays},\ }\href {https://doi.org/10.1103/PhysRevA.107.053508}
  {\bibfield  {journal} {\bibinfo  {journal} {Phys. Rev. A}\ }\textbf {\bibinfo
  {volume} {107}},\ \bibinfo {pages} {053508} (\bibinfo {year}
  {2023})}\BibitemShut {NoStop}%
\bibitem [{\citenamefont {Vu}(2022)}]{Floquet3}%
  \BibitemOpen
  \bibfield  {author} {\bibinfo {author} {\bibfnamefont {D.}~\bibnamefont
  {Vu}},\ }\bibfield  {title} {\bibinfo {title} {Dynamic bulk-boundary
  correspondence for anomalous floquet topology},\ }\href
  {https://doi.org/10.1103/PhysRevB.105.064304} {\bibfield  {journal} {\bibinfo
   {journal} {Phys. Rev. B}\ }\textbf {\bibinfo {volume} {105}},\ \bibinfo
  {pages} {064304} (\bibinfo {year} {2022})}\BibitemShut {NoStop}%
\bibitem [{\citenamefont {Li}\ \emph {et~al.}(2023)\citenamefont {Li},
  \citenamefont {Li}, \citenamefont {Yan}, \citenamefont {Li}, \citenamefont
  {Gong},\ and\ \citenamefont {Li}}]{Floquet4}%
  \BibitemOpen
  \bibfield  {author} {\bibinfo {author} {\bibfnamefont {M.}~\bibnamefont
  {Li}}, \bibinfo {author} {\bibfnamefont {C.}~\bibnamefont {Li}}, \bibinfo
  {author} {\bibfnamefont {L.}~\bibnamefont {Yan}}, \bibinfo {author}
  {\bibfnamefont {Q.}~\bibnamefont {Li}}, \bibinfo {author} {\bibfnamefont
  {Q.}~\bibnamefont {Gong}},\ and\ \bibinfo {author} {\bibfnamefont
  {Y.}~\bibnamefont {Li}},\ }\bibfield  {title} {\bibinfo {title} {Fractal
  photonic anomalous floquet topological insulators to generate multiple
  quantum chiral edge states},\ }\href@noop {} {\bibfield  {journal} {\bibinfo
  {journal} {Light: Science \& Applications}\ }\textbf {\bibinfo {volume}
  {12}},\ \bibinfo {pages} {262} (\bibinfo {year} {2023})}\BibitemShut
  {NoStop}%
\bibitem [{\citenamefont {Ren}\ \emph {et~al.}(2024)\citenamefont {Ren},
  \citenamefont {Yu}, \citenamefont {Wu}, \citenamefont {Qi}, \citenamefont
  {Wang}, \citenamefont {Yao}, \citenamefont {Ren}, \citenamefont {Guo},
  \citenamefont {Jiang}, \citenamefont {Chen} \emph {et~al.}}]{Anderson1}%
  \BibitemOpen
  \bibfield  {author} {\bibinfo {author} {\bibfnamefont {M.}~\bibnamefont
  {Ren}}, \bibinfo {author} {\bibfnamefont {Y.}~\bibnamefont {Yu}}, \bibinfo
  {author} {\bibfnamefont {B.}~\bibnamefont {Wu}}, \bibinfo {author}
  {\bibfnamefont {X.}~\bibnamefont {Qi}}, \bibinfo {author} {\bibfnamefont
  {Y.}~\bibnamefont {Wang}}, \bibinfo {author} {\bibfnamefont {X.}~\bibnamefont
  {Yao}}, \bibinfo {author} {\bibfnamefont {J.}~\bibnamefont {Ren}}, \bibinfo
  {author} {\bibfnamefont {Z.}~\bibnamefont {Guo}}, \bibinfo {author}
  {\bibfnamefont {H.}~\bibnamefont {Jiang}}, \bibinfo {author} {\bibfnamefont
  {H.}~\bibnamefont {Chen}}, \emph {et~al.},\ }\bibfield  {title} {\bibinfo
  {title} {Realization of gapped and ungapped photonic topological anderson
  insulators},\ }\href@noop {} {\bibfield  {journal} {\bibinfo  {journal}
  {Physical Review Letters}\ }\textbf {\bibinfo {volume} {132}},\ \bibinfo
  {pages} {066602} (\bibinfo {year} {2024})}\BibitemShut {NoStop}%
\bibitem [{\citenamefont {St{\"u}tzer}\ \emph {et~al.}(2018)\citenamefont
  {St{\"u}tzer}, \citenamefont {Plotnik}, \citenamefont {Lumer}, \citenamefont
  {Titum}, \citenamefont {Lindner}, \citenamefont {Segev}, \citenamefont
  {Rechtsman},\ and\ \citenamefont {Szameit}}]{Anderson2}%
  \BibitemOpen
  \bibfield  {author} {\bibinfo {author} {\bibfnamefont {S.}~\bibnamefont
  {St{\"u}tzer}}, \bibinfo {author} {\bibfnamefont {Y.}~\bibnamefont
  {Plotnik}}, \bibinfo {author} {\bibfnamefont {Y.}~\bibnamefont {Lumer}},
  \bibinfo {author} {\bibfnamefont {P.}~\bibnamefont {Titum}}, \bibinfo
  {author} {\bibfnamefont {N.~H.}\ \bibnamefont {Lindner}}, \bibinfo {author}
  {\bibfnamefont {M.}~\bibnamefont {Segev}}, \bibinfo {author} {\bibfnamefont
  {M.~C.}\ \bibnamefont {Rechtsman}},\ and\ \bibinfo {author} {\bibfnamefont
  {A.}~\bibnamefont {Szameit}},\ }\bibfield  {title} {\bibinfo {title}
  {Photonic topological anderson insulators},\ }\href@noop {} {\bibfield
  {journal} {\bibinfo  {journal} {Nature}\ }\textbf {\bibinfo {volume} {560}},\
  \bibinfo {pages} {461} (\bibinfo {year} {2018})}\BibitemShut {NoStop}%
\bibitem [{\citenamefont {Han}\ \emph {et~al.}(2023)\citenamefont {Han},
  \citenamefont {Chua}, \citenamefont {Zeng}, \citenamefont {Zhu},
  \citenamefont {Wang}, \citenamefont {Qiang}, \citenamefont {Jin},
  \citenamefont {Wang}, \citenamefont {Li}, \citenamefont {Davies} \emph
  {et~al.}}]{Majorana1}%
  \BibitemOpen
  \bibfield  {author} {\bibinfo {author} {\bibfnamefont {S.}~\bibnamefont
  {Han}}, \bibinfo {author} {\bibfnamefont {Y.}~\bibnamefont {Chua}}, \bibinfo
  {author} {\bibfnamefont {Y.}~\bibnamefont {Zeng}}, \bibinfo {author}
  {\bibfnamefont {B.}~\bibnamefont {Zhu}}, \bibinfo {author} {\bibfnamefont
  {C.}~\bibnamefont {Wang}}, \bibinfo {author} {\bibfnamefont {B.}~\bibnamefont
  {Qiang}}, \bibinfo {author} {\bibfnamefont {Y.}~\bibnamefont {Jin}}, \bibinfo
  {author} {\bibfnamefont {Q.}~\bibnamefont {Wang}}, \bibinfo {author}
  {\bibfnamefont {L.}~\bibnamefont {Li}}, \bibinfo {author} {\bibfnamefont
  {A.~G.}\ \bibnamefont {Davies}}, \emph {et~al.},\ }\bibfield  {title}
  {\bibinfo {title} {Photonic majorana quantum cascade laser with
  polarization-winding emission},\ }\href@noop {} {\bibfield  {journal}
  {\bibinfo  {journal} {Nature Communications}\ }\textbf {\bibinfo {volume}
  {14}},\ \bibinfo {pages} {707} (\bibinfo {year} {2023})}\BibitemShut
  {NoStop}%
\bibitem [{\citenamefont {Duggan}\ \emph {et~al.}(2020)\citenamefont {Duggan},
  \citenamefont {Mann},\ and\ \citenamefont {Al\`u}}]{non-reciprocity1}%
  \BibitemOpen
  \bibfield  {author} {\bibinfo {author} {\bibfnamefont {R.}~\bibnamefont
  {Duggan}}, \bibinfo {author} {\bibfnamefont {S.~A.}\ \bibnamefont {Mann}},\
  and\ \bibinfo {author} {\bibfnamefont {A.}~\bibnamefont {Al\`u}},\ }\bibfield
   {title} {\bibinfo {title} {Nonreciprocal photonic topological order driven
  by uniform optical pumping},\ }\href
  {https://doi.org/10.1103/PhysRevB.102.100303} {\bibfield  {journal} {\bibinfo
   {journal} {Phys. Rev. B}\ }\textbf {\bibinfo {volume} {102}},\ \bibinfo
  {pages} {100303} (\bibinfo {year} {2020})}\BibitemShut {NoStop}%
\bibitem [{\citenamefont {Rechtsman}(2023)}]{non-reciprocity2}%
  \BibitemOpen
  \bibfield  {author} {\bibinfo {author} {\bibfnamefont {M.~C.}\ \bibnamefont
  {Rechtsman}},\ }\bibfield  {title} {\bibinfo {title} {Reciprocal topological
  photonic crystals allow backscattering},\ }\href@noop {} {\bibfield
  {journal} {\bibinfo  {journal} {Nature Photonics}\ }\textbf {\bibinfo
  {volume} {17}},\ \bibinfo {pages} {383} (\bibinfo {year} {2023})}\BibitemShut
  {NoStop}%
\bibitem [{\citenamefont {Qian}\ \emph {et~al.}(2024)\citenamefont {Qian},
  \citenamefont {Zhang}, \citenamefont {Sun},\ and\ \citenamefont
  {Zhang}}]{non-Abelian1}%
  \BibitemOpen
  \bibfield  {author} {\bibinfo {author} {\bibfnamefont {L.}~\bibnamefont
  {Qian}}, \bibinfo {author} {\bibfnamefont {W.}~\bibnamefont {Zhang}},
  \bibinfo {author} {\bibfnamefont {H.}~\bibnamefont {Sun}},\ and\ \bibinfo
  {author} {\bibfnamefont {X.}~\bibnamefont {Zhang}},\ }\bibfield  {title}
  {\bibinfo {title} {Non-abelian topological bound states in the continuum},\
  }\href {https://doi.org/10.1103/PhysRevLett.132.046601} {\bibfield  {journal}
  {\bibinfo  {journal} {Phys. Rev. Lett.}\ }\textbf {\bibinfo {volume} {132}},\
  \bibinfo {pages} {046601} (\bibinfo {year} {2024})}\BibitemShut {NoStop}%
\bibitem [{\citenamefont {You}\ \emph {et~al.}(2022)\citenamefont {You},
  \citenamefont {Liang}, \citenamefont {Xie}, \citenamefont {Gao},
  \citenamefont {Ye}, \citenamefont {Zhu},\ and\ \citenamefont
  {Zhang}}]{non-Abelian2}%
  \BibitemOpen
  \bibfield  {author} {\bibinfo {author} {\bibfnamefont {O.}~\bibnamefont
  {You}}, \bibinfo {author} {\bibfnamefont {S.}~\bibnamefont {Liang}}, \bibinfo
  {author} {\bibfnamefont {B.}~\bibnamefont {Xie}}, \bibinfo {author}
  {\bibfnamefont {W.}~\bibnamefont {Gao}}, \bibinfo {author} {\bibfnamefont
  {W.}~\bibnamefont {Ye}}, \bibinfo {author} {\bibfnamefont {J.}~\bibnamefont
  {Zhu}},\ and\ \bibinfo {author} {\bibfnamefont {S.}~\bibnamefont {Zhang}},\
  }\bibfield  {title} {\bibinfo {title} {Observation of non-abelian thouless
  pump},\ }\href {https://doi.org/10.1103/PhysRevLett.128.244302} {\bibfield
  {journal} {\bibinfo  {journal} {Phys. Rev. Lett.}\ }\textbf {\bibinfo
  {volume} {128}},\ \bibinfo {pages} {244302} (\bibinfo {year}
  {2022})}\BibitemShut {NoStop}%
\bibitem [{\citenamefont {Parto}\ \emph {et~al.}(2023)\citenamefont {Parto},
  \citenamefont {Leefmans}, \citenamefont {Williams}, \citenamefont {Nori},\
  and\ \citenamefont {Marandi}}]{non-Abelian3}%
  \BibitemOpen
  \bibfield  {author} {\bibinfo {author} {\bibfnamefont {M.}~\bibnamefont
  {Parto}}, \bibinfo {author} {\bibfnamefont {C.}~\bibnamefont {Leefmans}},
  \bibinfo {author} {\bibfnamefont {J.}~\bibnamefont {Williams}}, \bibinfo
  {author} {\bibfnamefont {F.}~\bibnamefont {Nori}},\ and\ \bibinfo {author}
  {\bibfnamefont {A.}~\bibnamefont {Marandi}},\ }\bibfield  {title} {\bibinfo
  {title} {Non-abelian effects in dissipative photonic topological lattices},\
  }\href@noop {} {\bibfield  {journal} {\bibinfo  {journal} {Nature
  Communications}\ }\textbf {\bibinfo {volume} {14}},\ \bibinfo {pages} {1440}
  (\bibinfo {year} {2023})}\BibitemShut {NoStop}%
\bibitem [{\citenamefont {Huang}\ \emph {et~al.}(2024)\citenamefont {Huang},
  \citenamefont {He}, \citenamefont {Zhang}, \citenamefont {Zhang},
  \citenamefont {Liu}, \citenamefont {Feng}, \citenamefont {Liu}, \citenamefont
  {Cui}, \citenamefont {Huang}, \citenamefont {Zhang} \emph
  {et~al.}}]{hyperbolic1}%
  \BibitemOpen
  \bibfield  {author} {\bibinfo {author} {\bibfnamefont {L.}~\bibnamefont
  {Huang}}, \bibinfo {author} {\bibfnamefont {L.}~\bibnamefont {He}}, \bibinfo
  {author} {\bibfnamefont {W.}~\bibnamefont {Zhang}}, \bibinfo {author}
  {\bibfnamefont {H.}~\bibnamefont {Zhang}}, \bibinfo {author} {\bibfnamefont
  {D.}~\bibnamefont {Liu}}, \bibinfo {author} {\bibfnamefont {X.}~\bibnamefont
  {Feng}}, \bibinfo {author} {\bibfnamefont {F.}~\bibnamefont {Liu}}, \bibinfo
  {author} {\bibfnamefont {K.}~\bibnamefont {Cui}}, \bibinfo {author}
  {\bibfnamefont {Y.}~\bibnamefont {Huang}}, \bibinfo {author} {\bibfnamefont
  {W.}~\bibnamefont {Zhang}}, \emph {et~al.},\ }\bibfield  {title} {\bibinfo
  {title} {Hyperbolic photonic topological insulators},\ }\href@noop {}
  {\bibfield  {journal} {\bibinfo  {journal} {Nature Communications}\ }\textbf
  {\bibinfo {volume} {15}},\ \bibinfo {pages} {1647} (\bibinfo {year}
  {2024})}\BibitemShut {NoStop}%
\bibitem [{\citenamefont {Lin}\ \emph {et~al.}(2023)\citenamefont {Lin},
  \citenamefont {Wang}, \citenamefont {Liu}, \citenamefont {Xue}, \citenamefont
  {Zhang}, \citenamefont {Chong},\ and\ \citenamefont {Jiang}}]{PPCCM1}%
  \BibitemOpen
  \bibfield  {author} {\bibinfo {author} {\bibfnamefont {Z.-K.}\ \bibnamefont
  {Lin}}, \bibinfo {author} {\bibfnamefont {Q.}~\bibnamefont {Wang}}, \bibinfo
  {author} {\bibfnamefont {Y.}~\bibnamefont {Liu}}, \bibinfo {author}
  {\bibfnamefont {H.}~\bibnamefont {Xue}}, \bibinfo {author} {\bibfnamefont
  {B.}~\bibnamefont {Zhang}}, \bibinfo {author} {\bibfnamefont
  {Y.}~\bibnamefont {Chong}},\ and\ \bibinfo {author} {\bibfnamefont {J.-H.}\
  \bibnamefont {Jiang}},\ }\bibfield  {title} {\bibinfo {title} {Topological
  phenomena at defects in acoustic, photonic and solid-state lattices},\
  }\href@noop {} {\bibfield  {journal} {\bibinfo  {journal} {Nature Reviews
  Physics}\ }\textbf {\bibinfo {volume} {5}},\ \bibinfo {pages} {483} (\bibinfo
  {year} {2023})}\BibitemShut {NoStop}%
\bibitem [{\citenamefont {Zhang}\ \emph {et~al.}(2024)\citenamefont {Zhang},
  \citenamefont {Gu}, \citenamefont {Si},\ and\ \citenamefont {Ding}}]{PPCCM2}%
  \BibitemOpen
  \bibfield  {author} {\bibinfo {author} {\bibfnamefont {H.}~\bibnamefont
  {Zhang}}, \bibinfo {author} {\bibfnamefont {Z.}~\bibnamefont {Gu}}, \bibinfo
  {author} {\bibfnamefont {L.}~\bibnamefont {Si}},\ and\ \bibinfo {author}
  {\bibfnamefont {J.}~\bibnamefont {Ding}},\ }\bibfield  {title} {\bibinfo
  {title} {Realization of terahertz frequency selecting based on topological
  edge states with kagome photonic crystals},\ }\href
  {https://api.semanticscholar.org/CorpusID:267448996} {\bibfield  {journal}
  {\bibinfo  {journal} {Results in Physics}\ } (\bibinfo {year}
  {2024})}\BibitemShut {NoStop}%
\bibitem [{\citenamefont {Lin}\ \emph {et~al.}(2024)\citenamefont {Lin},
  \citenamefont {Chien},\ and\ \citenamefont {Hsueh}}]{PPCCM3}%
  \BibitemOpen
  \bibfield  {author} {\bibinfo {author} {\bibfnamefont {Y.-C.}\ \bibnamefont
  {Lin}}, \bibinfo {author} {\bibfnamefont {Y.-C.}\ \bibnamefont {Chien}},\
  and\ \bibinfo {author} {\bibfnamefont {W.-J.}\ \bibnamefont {Hsueh}},\
  }\bibfield  {title} {\bibinfo {title} {Topological slow-light in
  one-dimensional conjugated photonic systems},\ }\href
  {https://api.semanticscholar.org/CorpusID:267631968} {\bibfield  {journal}
  {\bibinfo  {journal} {Optics Communications}\ } (\bibinfo {year}
  {2024})}\BibitemShut {NoStop}%
\bibitem [{\citenamefont {Fan}\ \emph {et~al.}(2024)\citenamefont {Fan},
  \citenamefont {Chen}, \citenamefont {Zhu},\ and\ \citenamefont
  {Su}}]{PPCCM4}%
  \BibitemOpen
  \bibfield  {author} {\bibinfo {author} {\bibfnamefont {L.}~\bibnamefont
  {Fan}}, \bibinfo {author} {\bibfnamefont {Y.}~\bibnamefont {Chen}}, \bibinfo
  {author} {\bibfnamefont {J.}~\bibnamefont {Zhu}},\ and\ \bibinfo {author}
  {\bibfnamefont {Z.}~\bibnamefont {Su}},\ }\bibfield  {title} {\bibinfo
  {title} {Multi-band topological valley modes of flexural waves in
  micro-perforated phononic plates},\ }\href
  {https://doi.org/https://doi.org/10.1016/j.ijmecsci.2023.108916} {\bibfield
  {journal} {\bibinfo  {journal} {International Journal of Mechanical
  Sciences}\ }\textbf {\bibinfo {volume} {266}},\ \bibinfo {pages} {108916}
  (\bibinfo {year} {2024})}\BibitemShut {NoStop}%
\bibitem [{\citenamefont {Zou}\ \emph {et~al.}(2021)\citenamefont {Zou},
  \citenamefont {Chen}, \citenamefont {He}, \citenamefont {Bao}, \citenamefont
  {Lee}, \citenamefont {Sun},\ and\ \citenamefont {Zhang}}]{PPCCM5}%
  \BibitemOpen
  \bibfield  {author} {\bibinfo {author} {\bibfnamefont {D.}~\bibnamefont
  {Zou}}, \bibinfo {author} {\bibfnamefont {T.}~\bibnamefont {Chen}}, \bibinfo
  {author} {\bibfnamefont {W.}~\bibnamefont {He}}, \bibinfo {author}
  {\bibfnamefont {J.}~\bibnamefont {Bao}}, \bibinfo {author} {\bibfnamefont
  {C.~H.}\ \bibnamefont {Lee}}, \bibinfo {author} {\bibfnamefont
  {H.}~\bibnamefont {Sun}},\ and\ \bibinfo {author} {\bibfnamefont
  {X.}~\bibnamefont {Zhang}},\ }\bibfield  {title} {\bibinfo {title}
  {Observation of hybrid higher-order skin-topological effect in non-hermitian
  topolectrical circuits},\ }\href@noop {} {\bibfield  {journal} {\bibinfo
  {journal} {Nature Communications}\ }\textbf {\bibinfo {volume} {12}},\
  \bibinfo {pages} {7201} (\bibinfo {year} {2021})}\BibitemShut {NoStop}%
\bibitem [{\citenamefont {Gonz\'alez-Cuadra}\ \emph {et~al.}(2020)\citenamefont
  {Gonz\'alez-Cuadra}, \citenamefont {Dauphin}, \citenamefont {Grzybowski},
  \citenamefont {Lewenstein},\ and\ \citenamefont {Bermudez}}]{PPCCM6}%
  \BibitemOpen
  \bibfield  {author} {\bibinfo {author} {\bibfnamefont {D.}~\bibnamefont
  {Gonz\'alez-Cuadra}}, \bibinfo {author} {\bibfnamefont {A.}~\bibnamefont
  {Dauphin}}, \bibinfo {author} {\bibfnamefont {P.~R.}\ \bibnamefont
  {Grzybowski}}, \bibinfo {author} {\bibfnamefont {M.}~\bibnamefont
  {Lewenstein}},\ and\ \bibinfo {author} {\bibfnamefont {A.}~\bibnamefont
  {Bermudez}},\ }\bibfield  {title} {\bibinfo {title} {Dynamical solitons and
  boson fractionalization in cold-atom topological insulators},\ }\href
  {https://doi.org/10.1103/PhysRevLett.125.265301} {\bibfield  {journal}
  {\bibinfo  {journal} {Phys. Rev. Lett.}\ }\textbf {\bibinfo {volume} {125}},\
  \bibinfo {pages} {265301} (\bibinfo {year} {2020})}\BibitemShut {NoStop}%
\bibitem [{\citenamefont {Mugel}\ \emph {et~al.}(2016)\citenamefont {Mugel},
  \citenamefont {Celi}, \citenamefont {Massignan}, \citenamefont {Asb\'oth},
  \citenamefont {Lewenstein},\ and\ \citenamefont {Lobo}}]{PPCCM7}%
  \BibitemOpen
  \bibfield  {author} {\bibinfo {author} {\bibfnamefont {S.}~\bibnamefont
  {Mugel}}, \bibinfo {author} {\bibfnamefont {A.}~\bibnamefont {Celi}},
  \bibinfo {author} {\bibfnamefont {P.}~\bibnamefont {Massignan}}, \bibinfo
  {author} {\bibfnamefont {J.~K.}\ \bibnamefont {Asb\'oth}}, \bibinfo {author}
  {\bibfnamefont {M.}~\bibnamefont {Lewenstein}},\ and\ \bibinfo {author}
  {\bibfnamefont {C.}~\bibnamefont {Lobo}},\ }\bibfield  {title} {\bibinfo
  {title} {Topological bound states of a quantum walk with cold atoms},\ }\href
  {https://doi.org/10.1103/PhysRevA.94.023631} {\bibfield  {journal} {\bibinfo
  {journal} {Phys. Rev. A}\ }\textbf {\bibinfo {volume} {94}},\ \bibinfo
  {pages} {023631} (\bibinfo {year} {2016})}\BibitemShut {NoStop}%
\bibitem [{\citenamefont {Gao}\ \emph {et~al.}(2024)\citenamefont {Gao},
  \citenamefont {Zhang}, \citenamefont {Li}, \citenamefont {Zhang},
  \citenamefont {Chen}, \citenamefont {Du}, \citenamefont {Hou}, \citenamefont
  {Gu}, \citenamefont {Lun}, \citenamefont {Zhao}, \citenamefont {Zhao},
  \citenamefont {Qu}, \citenamefont {Jin}, \citenamefont {Wang}, \citenamefont
  {Chen}, \citenamefont {Liu}, \citenamefont {Huang}, \citenamefont {Gao},
  \citenamefont {Mostovoy}, \citenamefont {Hong}, \citenamefont {Cheong},\ and\
  \citenamefont {Wang}}]{PPCCM8}%
  \BibitemOpen
  \bibfield  {author} {\bibinfo {author} {\bibfnamefont {Z.}~\bibnamefont
  {Gao}}, \bibinfo {author} {\bibfnamefont {Y.}~\bibnamefont {Zhang}}, \bibinfo
  {author} {\bibfnamefont {X.}~\bibnamefont {Li}}, \bibinfo {author}
  {\bibfnamefont {X.}~\bibnamefont {Zhang}}, \bibinfo {author} {\bibfnamefont
  {X.}~\bibnamefont {Chen}}, \bibinfo {author} {\bibfnamefont {G.}~\bibnamefont
  {Du}}, \bibinfo {author} {\bibfnamefont {F.}~\bibnamefont {Hou}}, \bibinfo
  {author} {\bibfnamefont {B.}~\bibnamefont {Gu}}, \bibinfo {author}
  {\bibfnamefont {Y.}~\bibnamefont {Lun}}, \bibinfo {author} {\bibfnamefont
  {Y.}~\bibnamefont {Zhao}}, \bibinfo {author} {\bibfnamefont {Y.}~\bibnamefont
  {Zhao}}, \bibinfo {author} {\bibfnamefont {Z.}~\bibnamefont {Qu}}, \bibinfo
  {author} {\bibfnamefont {K.}~\bibnamefont {Jin}}, \bibinfo {author}
  {\bibfnamefont {X.}~\bibnamefont {Wang}}, \bibinfo {author} {\bibfnamefont
  {Y.}~\bibnamefont {Chen}}, \bibinfo {author} {\bibfnamefont {Z.}~\bibnamefont
  {Liu}}, \bibinfo {author} {\bibfnamefont {H.}~\bibnamefont {Huang}}, \bibinfo
  {author} {\bibfnamefont {P.}~\bibnamefont {Gao}}, \bibinfo {author}
  {\bibfnamefont {M.}~\bibnamefont {Mostovoy}}, \bibinfo {author}
  {\bibfnamefont {J.}~\bibnamefont {Hong}}, \bibinfo {author} {\bibfnamefont
  {S.-W.}\ \bibnamefont {Cheong}},\ and\ \bibinfo {author} {\bibfnamefont
  {X.}~\bibnamefont {Wang}},\ }\bibfield  {title} {\bibinfo {title} {Mechanical
  manipulation for ordered topological defects},\ }\href
  {https://doi.org/10.1126/sciadv.adi5894} {\bibfield  {journal} {\bibinfo
  {journal} {Science Advances}\ }\textbf {\bibinfo {volume} {10}},\ \bibinfo
  {pages} {eadi5894} (\bibinfo {year} {2024})},\ \Eprint
  {https://arxiv.org/abs/https://www.science.org/doi/pdf/10.1126/sciadv.adi5894}
  {https://www.science.org/doi/pdf/10.1126/sciadv.adi5894} \BibitemShut
  {NoStop}%
\bibitem [{\citenamefont {Li}\ \emph {et~al.}(2020)\citenamefont {Li},
  \citenamefont {Zhirihin}, \citenamefont {Gorlach}, \citenamefont {Ni},
  \citenamefont {Filonov}, \citenamefont {Slobozhanyuk}, \citenamefont
  {Al{\`u}},\ and\ \citenamefont {Khanikaev}}]{HO1}%
  \BibitemOpen
  \bibfield  {author} {\bibinfo {author} {\bibfnamefont {M.}~\bibnamefont
  {Li}}, \bibinfo {author} {\bibfnamefont {D.}~\bibnamefont {Zhirihin}},
  \bibinfo {author} {\bibfnamefont {M.}~\bibnamefont {Gorlach}}, \bibinfo
  {author} {\bibfnamefont {X.}~\bibnamefont {Ni}}, \bibinfo {author}
  {\bibfnamefont {D.}~\bibnamefont {Filonov}}, \bibinfo {author} {\bibfnamefont
  {A.}~\bibnamefont {Slobozhanyuk}}, \bibinfo {author} {\bibfnamefont
  {A.}~\bibnamefont {Al{\`u}}},\ and\ \bibinfo {author} {\bibfnamefont {A.~B.}\
  \bibnamefont {Khanikaev}},\ }\bibfield  {title} {\bibinfo {title}
  {Higher-order topological states in photonic kagome crystals with long-range
  interactions},\ }\href@noop {} {\bibfield  {journal} {\bibinfo  {journal}
  {Nature Photonics}\ }\textbf {\bibinfo {volume} {14}},\ \bibinfo {pages} {89}
  (\bibinfo {year} {2020})}\BibitemShut {NoStop}%
\bibitem [{\citenamefont {Rao}\ \emph {et~al.}(2024)\citenamefont {Rao},
  \citenamefont {Shi}, \citenamefont {Rao}, \citenamefont {Yang}, \citenamefont
  {Song}, \citenamefont {Chen}, \citenamefont {Dong}, \citenamefont {Yu},\ and\
  \citenamefont {Yu}}]{HO2}%
  \BibitemOpen
  \bibfield  {author} {\bibinfo {author} {\bibfnamefont {M.}~\bibnamefont
  {Rao}}, \bibinfo {author} {\bibfnamefont {F.}~\bibnamefont {Shi}}, \bibinfo
  {author} {\bibfnamefont {Z.}~\bibnamefont {Rao}}, \bibinfo {author}
  {\bibfnamefont {J.}~\bibnamefont {Yang}}, \bibinfo {author} {\bibfnamefont
  {C.}~\bibnamefont {Song}}, \bibinfo {author} {\bibfnamefont {X.}~\bibnamefont
  {Chen}}, \bibinfo {author} {\bibfnamefont {J.}~\bibnamefont {Dong}}, \bibinfo
  {author} {\bibfnamefont {Y.}~\bibnamefont {Yu}},\ and\ \bibinfo {author}
  {\bibfnamefont {S.}~\bibnamefont {Yu}},\ }\bibfield  {title} {\bibinfo
  {title} {Single photon emitter deterministically coupled to a topological
  corner state},\ }\href@noop {} {\bibfield  {journal} {\bibinfo  {journal}
  {Light: Science \& Applications}\ }\textbf {\bibinfo {volume} {13}},\
  \bibinfo {pages} {19} (\bibinfo {year} {2024})}\BibitemShut {NoStop}%
\bibitem [{\citenamefont {Mittal}\ \emph {et~al.}(2019)\citenamefont {Mittal},
  \citenamefont {Orre}, \citenamefont {Zhu}, \citenamefont {Gorlach},
  \citenamefont {Poddubny},\ and\ \citenamefont {Hafezi}}]{HO3}%
  \BibitemOpen
  \bibfield  {author} {\bibinfo {author} {\bibfnamefont {S.}~\bibnamefont
  {Mittal}}, \bibinfo {author} {\bibfnamefont {V.~V.}\ \bibnamefont {Orre}},
  \bibinfo {author} {\bibfnamefont {G.}~\bibnamefont {Zhu}}, \bibinfo {author}
  {\bibfnamefont {M.~A.}\ \bibnamefont {Gorlach}}, \bibinfo {author}
  {\bibfnamefont {A.}~\bibnamefont {Poddubny}},\ and\ \bibinfo {author}
  {\bibfnamefont {M.}~\bibnamefont {Hafezi}},\ }\bibfield  {title} {\bibinfo
  {title} {Photonic quadrupole topological phases},\ }\href@noop {} {\bibfield
  {journal} {\bibinfo  {journal} {Nature Photonics}\ }\textbf {\bibinfo
  {volume} {13}},\ \bibinfo {pages} {692} (\bibinfo {year} {2019})}\BibitemShut
  {NoStop}%
\bibitem [{\citenamefont {Xie}\ \emph {et~al.}(2018{\natexlab{a}})\citenamefont
  {Xie}, \citenamefont {Wang}, \citenamefont {Wang}, \citenamefont {Zhu},
  \citenamefont {Jiang}, \citenamefont {Lu},\ and\ \citenamefont {Chen}}]{HO4}%
  \BibitemOpen
  \bibfield  {author} {\bibinfo {author} {\bibfnamefont {B.-Y.}\ \bibnamefont
  {Xie}}, \bibinfo {author} {\bibfnamefont {H.-F.}\ \bibnamefont {Wang}},
  \bibinfo {author} {\bibfnamefont {H.-X.}\ \bibnamefont {Wang}}, \bibinfo
  {author} {\bibfnamefont {X.-Y.}\ \bibnamefont {Zhu}}, \bibinfo {author}
  {\bibfnamefont {J.-H.}\ \bibnamefont {Jiang}}, \bibinfo {author}
  {\bibfnamefont {M.-H.}\ \bibnamefont {Lu}},\ and\ \bibinfo {author}
  {\bibfnamefont {Y.-F.}\ \bibnamefont {Chen}},\ }\bibfield  {title} {\bibinfo
  {title} {Second-order photonic topological insulator with corner states},\
  }\href {https://doi.org/10.1103/PhysRevB.98.205147} {\bibfield  {journal}
  {\bibinfo  {journal} {Phys. Rev. B}\ }\textbf {\bibinfo {volume} {98}},\
  \bibinfo {pages} {205147} (\bibinfo {year} {2018}{\natexlab{a}})}\BibitemShut
  {NoStop}%
\bibitem [{\citenamefont {He}\ \emph {et~al.}(2020)\citenamefont {He},
  \citenamefont {Addison}, \citenamefont {Mele},\ and\ \citenamefont
  {Zhen}}]{HO5}%
  \BibitemOpen
  \bibfield  {author} {\bibinfo {author} {\bibfnamefont {L.}~\bibnamefont
  {He}}, \bibinfo {author} {\bibfnamefont {Z.}~\bibnamefont {Addison}},
  \bibinfo {author} {\bibfnamefont {E.~J.}\ \bibnamefont {Mele}},\ and\
  \bibinfo {author} {\bibfnamefont {B.}~\bibnamefont {Zhen}},\ }\bibfield
  {title} {\bibinfo {title} {Quadrupole topological photonic crystals},\
  }\href@noop {} {\bibfield  {journal} {\bibinfo  {journal} {Nature
  communications}\ }\textbf {\bibinfo {volume} {11}},\ \bibinfo {pages} {3119}
  (\bibinfo {year} {2020})}\BibitemShut {NoStop}%
\bibitem [{\citenamefont {Xie}\ \emph {et~al.}(2018{\natexlab{b}})\citenamefont
  {Xie}, \citenamefont {Su}, \citenamefont {Wang}, \citenamefont {Su},
  \citenamefont {Shen}, \citenamefont {Zhan}, \citenamefont {Lu}, \citenamefont
  {Wang},\ and\ \citenamefont {Chen}}]{HO6}%
  \BibitemOpen
  \bibfield  {author} {\bibinfo {author} {\bibfnamefont {B.}~\bibnamefont
  {Xie}}, \bibinfo {author} {\bibfnamefont {G.}~\bibnamefont {Su}}, \bibinfo
  {author} {\bibfnamefont {H.}~\bibnamefont {Wang}}, \bibinfo {author}
  {\bibfnamefont {H.}~\bibnamefont {Su}}, \bibinfo {author} {\bibfnamefont
  {X.}~\bibnamefont {Shen}}, \bibinfo {author} {\bibfnamefont {P.}~\bibnamefont
  {Zhan}}, \bibinfo {author} {\bibfnamefont {M.-H.}\ \bibnamefont {Lu}},
  \bibinfo {author} {\bibfnamefont {Z.}~\bibnamefont {Wang}},\ and\ \bibinfo
  {author} {\bibfnamefont {Y.-F.}\ \bibnamefont {Chen}},\ }\bibfield  {title}
  {\bibinfo {title} {Visualization of higher-order topological insulating
  phases in two-dimensional dielectric photonic crystals.},\ }\href
  {https://api.semanticscholar.org/CorpusID:118929249} {\bibfield  {journal}
  {\bibinfo  {journal} {Physical review letters}\ }\textbf {\bibinfo {volume}
  {122 23}},\ \bibinfo {pages} {233903} (\bibinfo {year}
  {2018}{\natexlab{b}})}\BibitemShut {NoStop}%
\bibitem [{\citenamefont {Guo}\ \emph {et~al.}(2021)\citenamefont {Guo},
  \citenamefont {Jiang}, \citenamefont {Jiang}, \citenamefont {Ren},\ and\
  \citenamefont {Chen}}]{SR1}%
  \BibitemOpen
  \bibfield  {author} {\bibinfo {author} {\bibfnamefont {Z.}~\bibnamefont
  {Guo}}, \bibinfo {author} {\bibfnamefont {J.}~\bibnamefont {Jiang}}, \bibinfo
  {author} {\bibfnamefont {H.}~\bibnamefont {Jiang}}, \bibinfo {author}
  {\bibfnamefont {J.}~\bibnamefont {Ren}},\ and\ \bibinfo {author}
  {\bibfnamefont {H.}~\bibnamefont {Chen}},\ }\bibfield  {title} {\bibinfo
  {title} {Observation of topological bound states in a double
  su-schrieffer-heeger chain composed of split ring resonators},\ }\href
  {https://doi.org/10.1103/PhysRevResearch.3.013122} {\bibfield  {journal}
  {\bibinfo  {journal} {Phys. Rev. Res.}\ }\textbf {\bibinfo {volume} {3}},\
  \bibinfo {pages} {013122} (\bibinfo {year} {2021})}\BibitemShut {NoStop}%
\bibitem [{\citenamefont {Chen}\ \emph {et~al.}(2022)\citenamefont {Chen},
  \citenamefont {Chen}, \citenamefont {Ding}, \citenamefont {Xiang},
  \citenamefont {Mao},\ and\ \citenamefont {Zhu}}]{SR2}%
  \BibitemOpen
  \bibfield  {author} {\bibinfo {author} {\bibfnamefont {C.}~\bibnamefont
  {Chen}}, \bibinfo {author} {\bibfnamefont {T.}~\bibnamefont {Chen}}, \bibinfo
  {author} {\bibfnamefont {W.}~\bibnamefont {Ding}}, \bibinfo {author}
  {\bibfnamefont {X.}~\bibnamefont {Xiang}}, \bibinfo {author} {\bibfnamefont
  {F.}~\bibnamefont {Mao}},\ and\ \bibinfo {author} {\bibfnamefont
  {J.}~\bibnamefont {Zhu}},\ }\bibfield  {title} {\bibinfo {title} {Split-ring
  resonator coupling-induced tunable acoustic second-order topological
  insulators},\ }\href {https://doi.org/10.1103/PhysRevB.106.045403} {\bibfield
   {journal} {\bibinfo  {journal} {Phys. Rev. B}\ }\textbf {\bibinfo {volume}
  {106}},\ \bibinfo {pages} {045403} (\bibinfo {year} {2022})}\BibitemShut
  {NoStop}%
\bibitem [{\citenamefont {Jiang}\ \emph {et~al.}(2020)\citenamefont {Jiang},
  \citenamefont {Ren}, \citenamefont {Guo}, \citenamefont {Zhu}, \citenamefont
  {Long}, \citenamefont {Jiang},\ and\ \citenamefont {Chen}}]{SR3}%
  \BibitemOpen
  \bibfield  {author} {\bibinfo {author} {\bibfnamefont {J.}~\bibnamefont
  {Jiang}}, \bibinfo {author} {\bibfnamefont {J.}~\bibnamefont {Ren}}, \bibinfo
  {author} {\bibfnamefont {Z.}~\bibnamefont {Guo}}, \bibinfo {author}
  {\bibfnamefont {W.}~\bibnamefont {Zhu}}, \bibinfo {author} {\bibfnamefont
  {Y.}~\bibnamefont {Long}}, \bibinfo {author} {\bibfnamefont {H.}~\bibnamefont
  {Jiang}},\ and\ \bibinfo {author} {\bibfnamefont {H.}~\bibnamefont {Chen}},\
  }\bibfield  {title} {\bibinfo {title} {Seeing topological winding number and
  band inversion in photonic dimer chain of split-ring resonators},\ }\href
  {https://doi.org/10.1103/PhysRevB.101.165427} {\bibfield  {journal} {\bibinfo
   {journal} {Phys. Rev. B}\ }\textbf {\bibinfo {volume} {101}},\ \bibinfo
  {pages} {165427} (\bibinfo {year} {2020})}\BibitemShut {NoStop}%
\bibitem [{\citenamefont {Mittal}\ \emph {et~al.}(2021)\citenamefont {Mittal},
  \citenamefont {Moille}, \citenamefont {Srinivasan}, \citenamefont {Chembo},\
  and\ \citenamefont {Hafezi}}]{SR4}%
  \BibitemOpen
  \bibfield  {author} {\bibinfo {author} {\bibfnamefont {S.}~\bibnamefont
  {Mittal}}, \bibinfo {author} {\bibfnamefont {G.}~\bibnamefont {Moille}},
  \bibinfo {author} {\bibfnamefont {K.}~\bibnamefont {Srinivasan}}, \bibinfo
  {author} {\bibfnamefont {Y.~K.}\ \bibnamefont {Chembo}},\ and\ \bibinfo
  {author} {\bibfnamefont {M.}~\bibnamefont {Hafezi}},\ }\bibfield  {title}
  {\bibinfo {title} {Topological frequency combs and nested temporal
  solitons},\ }\href@noop {} {\bibfield  {journal} {\bibinfo  {journal} {Nature
  Physics}\ }\textbf {\bibinfo {volume} {17}},\ \bibinfo {pages} {1169}
  (\bibinfo {year} {2021})}\BibitemShut {NoStop}%
\bibitem [{\citenamefont {Guo}\ \emph {et~al.}(2018)\citenamefont {Guo},
  \citenamefont {Jiang}, \citenamefont {Sun}, \citenamefont {Li},\ and\
  \citenamefont {Chen}}]{SR5}%
  \BibitemOpen
  \bibfield  {author} {\bibinfo {author} {\bibfnamefont {Z.}~\bibnamefont
  {Guo}}, \bibinfo {author} {\bibfnamefont {H.}~\bibnamefont {Jiang}}, \bibinfo
  {author} {\bibfnamefont {Y.}~\bibnamefont {Sun}}, \bibinfo {author}
  {\bibfnamefont {Y.}~\bibnamefont {Li}},\ and\ \bibinfo {author}
  {\bibfnamefont {H.}~\bibnamefont {Chen}},\ }\bibfield  {title} {\bibinfo
  {title} {Asymmetric topological edge states in a quasiperiodic harper chain
  composed of split-ring resonators},\ }\href
  {https://doi.org/10.1364/OL.43.005142} {\bibfield  {journal} {\bibinfo
  {journal} {Opt. Lett.}\ }\textbf {\bibinfo {volume} {43}},\ \bibinfo {pages}
  {5142} (\bibinfo {year} {2018})}\BibitemShut {NoStop}%
\bibitem [{\citenamefont {Ao}\ \emph {et~al.}(2020)\citenamefont {Ao},
  \citenamefont {Hu}, \citenamefont {You}, \citenamefont {Lu}, \citenamefont
  {Fu}, \citenamefont {Wang},\ and\ \citenamefont {Gong}}]{SR6}%
  \BibitemOpen
  \bibfield  {author} {\bibinfo {author} {\bibfnamefont {Y.}~\bibnamefont
  {Ao}}, \bibinfo {author} {\bibfnamefont {X.}~\bibnamefont {Hu}}, \bibinfo
  {author} {\bibfnamefont {Y.}~\bibnamefont {You}}, \bibinfo {author}
  {\bibfnamefont {C.}~\bibnamefont {Lu}}, \bibinfo {author} {\bibfnamefont
  {Y.}~\bibnamefont {Fu}}, \bibinfo {author} {\bibfnamefont {X.}~\bibnamefont
  {Wang}},\ and\ \bibinfo {author} {\bibfnamefont {Q.}~\bibnamefont {Gong}},\
  }\bibfield  {title} {\bibinfo {title} {Topological phase transition in the
  non-hermitian coupled resonator array},\ }\href
  {https://doi.org/10.1103/PhysRevLett.125.013902} {\bibfield  {journal}
  {\bibinfo  {journal} {Phys. Rev. Lett.}\ }\textbf {\bibinfo {volume} {125}},\
  \bibinfo {pages} {013902} (\bibinfo {year} {2020})}\BibitemShut {NoStop}%
\bibitem [{\citenamefont {Zhao}\ \emph {et~al.}(2018)\citenamefont {Zhao},
  \citenamefont {Miao}, \citenamefont {Teimourpour}, \citenamefont {Malzard},
  \citenamefont {El-Ganainy}, \citenamefont {Schomerus},\ and\ \citenamefont
  {Feng}}]{SR7}%
  \BibitemOpen
  \bibfield  {author} {\bibinfo {author} {\bibfnamefont {H.}~\bibnamefont
  {Zhao}}, \bibinfo {author} {\bibfnamefont {P.}~\bibnamefont {Miao}}, \bibinfo
  {author} {\bibfnamefont {M.~H.}\ \bibnamefont {Teimourpour}}, \bibinfo
  {author} {\bibfnamefont {S.}~\bibnamefont {Malzard}}, \bibinfo {author}
  {\bibfnamefont {R.}~\bibnamefont {El-Ganainy}}, \bibinfo {author}
  {\bibfnamefont {H.}~\bibnamefont {Schomerus}},\ and\ \bibinfo {author}
  {\bibfnamefont {L.}~\bibnamefont {Feng}},\ }\bibfield  {title} {\bibinfo
  {title} {Topological hybrid silicon microlasers},\ }\href@noop {} {\bibfield
  {journal} {\bibinfo  {journal} {Nature communications}\ }\textbf {\bibinfo
  {volume} {9}},\ \bibinfo {pages} {981} (\bibinfo {year} {2018})}\BibitemShut
  {NoStop}%
\bibitem [{\citenamefont {Lin}\ \emph {et~al.}(2021)\citenamefont {Lin},
  \citenamefont {Ke}, \citenamefont {Zhu},\ and\ \citenamefont {Li}}]{SR8}%
  \BibitemOpen
  \bibfield  {author} {\bibinfo {author} {\bibfnamefont {Z.}~\bibnamefont
  {Lin}}, \bibinfo {author} {\bibfnamefont {S.}~\bibnamefont {Ke}}, \bibinfo
  {author} {\bibfnamefont {X.}~\bibnamefont {Zhu}},\ and\ \bibinfo {author}
  {\bibfnamefont {X.}~\bibnamefont {Li}},\ }\bibfield  {title} {\bibinfo
  {title} {Square-root non-bloch topological insulators in non-hermitian ring
  resonators},\ }\href {https://doi.org/10.1364/OE.419852} {\bibfield
  {journal} {\bibinfo  {journal} {Opt. Express}\ }\textbf {\bibinfo {volume}
  {29}},\ \bibinfo {pages} {8462} (\bibinfo {year} {2021})}\BibitemShut
  {NoStop}%
\bibitem [{\citenamefont {Dutt}\ \emph {et~al.}(2020)\citenamefont {Dutt},
  \citenamefont {Minkov}, \citenamefont {Williamson},\ and\ \citenamefont
  {Fan}}]{SR9}%
  \BibitemOpen
  \bibfield  {author} {\bibinfo {author} {\bibfnamefont {A.}~\bibnamefont
  {Dutt}}, \bibinfo {author} {\bibfnamefont {M.}~\bibnamefont {Minkov}},
  \bibinfo {author} {\bibfnamefont {I.~A.}\ \bibnamefont {Williamson}},\ and\
  \bibinfo {author} {\bibfnamefont {S.}~\bibnamefont {Fan}},\ }\bibfield
  {title} {\bibinfo {title} {Higher-order topological insulators in synthetic
  dimensions},\ }\href@noop {} {\bibfield  {journal} {\bibinfo  {journal}
  {Light: Science \& Applications}\ }\textbf {\bibinfo {volume} {9}},\ \bibinfo
  {pages} {131} (\bibinfo {year} {2020})}\BibitemShut {NoStop}%
\bibitem [{\citenamefont {Lazarides}\ and\ \citenamefont
  {Tsironis}(2020)}]{SR10}%
  \BibitemOpen
  \bibfield  {author} {\bibinfo {author} {\bibfnamefont {N.}~\bibnamefont
  {Lazarides}}\ and\ \bibinfo {author} {\bibfnamefont {G.~P.}\ \bibnamefont
  {Tsironis}},\ }\bibfield  {title} {\bibinfo {title} {Topological split-ring
  resonator based metamaterials with $\mathcal{PT}$ symmetry relying on gain
  and loss},\ }\href {https://doi.org/10.1103/PhysRevB.102.064306} {\bibfield
  {journal} {\bibinfo  {journal} {Phys. Rev. B}\ }\textbf {\bibinfo {volume}
  {102}},\ \bibinfo {pages} {064306} (\bibinfo {year} {2020})}\BibitemShut
  {NoStop}%
\end{thebibliography}%
	
\end{document}